\documentclass[11pt,a4paper]{article}
\pdfoutput=1

\usepackage{geometry}
\usepackage[numbers,sort&compress]{natbib}
\usepackage{amsmath}
\usepackage{amssymb}
\usepackage[dvips]{graphicx}
\usepackage{pstricks}
\usepackage{bm}
\usepackage{pbox}
\usepackage{placeins}
\usepackage{graphicx}
\usepackage{caption}
\usepackage{subcaption}
\usepackage[T1]{fontenc}
\usepackage{footnote}
\usepackage{pdfpages}
\usepackage{hhline}
\usepackage{multirow}
\usepackage{enumitem}
\usepackage{slashed}
\usepackage[nottoc,notlot,notlof]{tocbibind}
\usepackage[title,titletoc]{appendix}
\usepackage{dsfont}
\allowdisplaybreaks
\usepackage{float}
\usepackage{bbold}
\usepackage{cases}
\usepackage{tikz}
\newcommand*\circled[1]{\tikz[baseline=(char.base)]{
            \node[shape=circle,draw,inner sep=2pt] (char) {#1};}}

\newcommand{\matrixel}[3]{\left< #1 \vphantom{#2#3} \right|
 #2 \left| #3 \vphantom{#1#2} \right>}

\definecolor{MyDarkBlue}{rgb}{0.1, 0.1, 0.8} 
\definecolor{MyLightBlue}{rgb}{0.22,0.51,0.9}
\definecolor{MyGreen}{rgb}{0.0, 0.5, 0.0}
\definecolor{BrickRed}{rgb}{0.8, 0.25, 0.33}
\usepackage[colorlinks=true, linkcolor=blue, citecolor=MyDarkBlue, urlcolor=MyLightBlue, bookmarksnumbered=true, bookmarksopen]{hyperref}
\hypersetup{colorlinks, citecolor=blue,linkcolor=red,  urlcolor=MyLightBlue}

\usepackage{pifont}
\newcommand{\cmark}{\ding{51}}%
\newcommand{\xmark}{\ding{55}}%

\begin{document}
\vspace*{-0.2in}
\begin{flushright}

\end{flushright}
\vspace{0.5cm}
\begin{center}
{\Large\bf 
Minimal $SU(5)$ GUTs with vectorlike fermions
}\\
\end{center}
\vspace{0.5cm}
\renewcommand{\thefootnote}{\fnsymbol{footnote}}
\begin{center}
{\large
{}~\textbf{Stefan Antusch}\footnote[1]{E-mail:  
\textcolor{MyLightBlue}{stefan.antusch@unibas.ch}},
{}~\textbf{Kevin Hinze}\footnote[4]{E-mail:  
\textcolor{MyLightBlue}{kevin.hinze@unibas.ch}}, and 
{}~\textbf{Shaikh Saad}\footnote[2]{E-mail:  
\textcolor{MyLightBlue}{shaikh.saad@unibas.ch}}
}
\vspace{0.5cm}

{\em Department of Physics, University of Basel,\\ Klingelbergstrasse\ 82, CH-4056 Basel, Switzerland}
\end{center}

\renewcommand{\thefootnote}{\arabic{footnote}}
\setcounter{footnote}{0}
\thispagestyle{empty}

\begin{abstract}
In this work, we attempt to answer the question, ``What is the minimal viable renormalizable $SU(5)$ GUT with representations no higher than adjoints?''. We find that an $SU(5)$ model with a pair of vectorlike fermions $5_F+\overline{5}_F$, as well as two copies of $15_H$ Higgs fields, is the minimal candidate that accommodates for correct charged fermion and neutrino masses and can also address the matter-antimatter asymmetry of the universe. Our results show that the presented model is highly predictive and will be fully tested by a combination of upcoming proton decay experiments, collider searches, and low-energy experiments in search of flavor violations. Moreover, we also entertain the possibility of adding a pair of vectorlike fermions $10_F+\overline{10}_F$ or $15_F+\overline{15}_F$ (instead of a $5_F+\overline{5}_F$). Our study reveals that the entire parameter space of these two models, even with minimal particle content, cannot be fully probed due to a possible longer proton lifetime beyond the reach of Hyper-Kamiokande.   
\end{abstract}
\newpage
{\hypersetup{linkcolor=black}
\tableofcontents}
\setcounter{footnote}{0}

\setcounter{page}{1} 
\section{Introduction}
The minimal simple group containing the entire gauge group of the Standard Model (SM)  is $SU(5)$, which has the same rank as the SM group. The minimal grand unified theory (GUT)~\cite{Pati:1973rp,Pati:1974yy, Georgi:1974sy, Georgi:1974yf, Georgi:1974my, Fritzsch:1974nn}  based on $SU(5)$ gauge symmetry, namely, the Georgi–Glashow (GG) model~\cite{Georgi:1974sy}, embeds all SM fermions of a single generation into one $\overline 5_F$ and one $10_F$ dimensional representation. The scalar sector of this theory is also exceedingly simple, consisting only of a fundamental $5_H$ and an adjoint $24_H$ Higgs. Despite its simplicity, the GG model suffers from fatal flaws, such as (i) it predicts a wrong mass relation between the down-type quarks and the charged leptons, (ii) gauge coupling unification does not take place, and (iii) the neutrinos remain massless.

There are several ways to overcome the drawbacks of the GG model, e.g., extending the particle content by a $45_H$~\cite{Georgi:1979df} dimensional Higgs representation can cure~\cite{Dorsner:2006dj} the first two problems listed above. However, neutrinos still remain massless. Straightforward ways to give neutrinos a nonzero mass are the implementation of a (a) type-I seesaw~\cite{Minkowski:1977sc,Yanagida:1979as,Glashow:1979nm,Gell-Mann:1979vob,Mohapatra:1979ia}, (b) type-II seesaw~\cite{Magg:1980ut,Schechter:1980gr,Lazarides:1980nt,Mohapatra:1980yp}, or (c) type-III seesaw~\cite{Foot:1988aq} mechanism.  The first of these possibilities requires the addition of at least two gauge-singlet right-chiral  neutrinos~\cite{Antusch:2021yqe}, while the second (third) option can be achieved by introducing a scalar (fermion) in the $15_H$~\cite{Dorsner:2005fq,Antusch:2022afk,Calibbi:2022wko} ($24_F$~\cite{Bajc:2006ia,Antusch:2023kli}) dimensional representation.

An alternative to these tree-level neutrino mass  mechanisms is to generate it via quantum corrections.  The most economical choice for this possibility utilizing smaller dimensional representations is to generate neutrino mass at the one-loop level by extending the GG model with a scalar $35_H$ and a vectorlike fermion in the $15_F+\overline{15}_F$~\cite{Dorsner:2019vgf,Dorsner:2021qwg,Antusch:2023jok} representation. If a $45_H$ Higgs is used instead, neutrino masses at one-loop can arise by adding a scalar in the $10_H$ representation~\cite{Wolfenstein:1980sf,Barbieri:1981yw,Perez:2016qbo}. A realization of a two-loop neutrino mass model, however, requires a non-minimal particle content, see for example, Ref.~\cite{Saad:2019vjo}.

\begin{table}[b!]\centering
        \begin{tabular}{|c|c|c|c|}\hline
            $Y_e\neq Y_d$ & $M_\nu\neq 0$ & Leptogenesis & Proton lifetime (years)
            \\\hline\hline
             & $2\times 1_F$ & \color{teal}\cmark  & \color{red}{$10^{28}$}
            \\\cline{2-4}
             & $2\times 24_F$ & \color{teal}\cmark\color{black}/\color{red}\xmark & \color{red}{$10^{33}$}\color{black}/\color{teal}{$10^{39}$} \\\cline{2-4}
            \raisebox{2ex}[0pt]{$5_F+\overline{5}_F$} & $1\times 15_H$ & \color{red}\xmark & \color{red}{$10^{31}$}
            \\\cline{2-4}
             & $2\times 15_H$ & \color{teal}\cmark & \color{teal}{$10^{35}$} 
            \\\hline\hline
            $10_F+\overline{10}_F/$ &&& \\
            $15_F+\overline{15}_F$ & \raisebox{2ex}[0pt]{$2\times 1_F$} & \raisebox{2ex}[0pt]{\color{teal}\cmark} & \raisebox{2ex}[0pt]{\color{teal}{$10^{38}$}}
            \\\hline
        \end{tabular}
        \caption{Various $SU(5)$ GUT scenarios in which the wrong mass relation between down-type quarks and charged leptons is corrected by the introduction of a single pair of VLFs, while neutrino masses are generated by one of the seesaw mechanisms. The third column indicates whether leptogenesis can be realized in a given scenario, whereas the forth column indicates the maximal proton lifetime for each scenario. For the case of $5_F+\overline{5}_F$ VLFs with the type-III seesaw, successful implementation of leptogenesis is not viable due to too rapid proton decay. On the contrary, a large GUT scale can be obtained in this scenario if leptogenesis constraints are not imposed. See text for details.}
        \label{tab:max_proton_lifetime_models}
\end{table}

In this work, we aim to answer to the question, ``What is the minimal viable renormalizable $SU(5)$ GUT with representations no higher than adjoints?''. In this context, within a renormalizable framework, the only  way to correct the aforementioned wrong mass relation is to introduce a pair of VLFs: \circled{1} $5_F+\overline{5}_F$, \circled{2} $10_F+\overline{10}_F$, or \circled{3} $15_F+\overline{15}_F$. For the first case with $5_F+\overline{5}_F$, the type-I seesaw mechanism to generate the observed neutrino masses is not viable since gauge couplings unify at such a low scale that it is ruled out by proton decay experiments. Our analysis shows that a type-II seesaw with a single $15_H$ Higgs is also not feasible for the same reason. Therefore, we study a scenario with two copies of $15_H$ Higgs and find that the proposed model has high predictive power and will be tested by the upcoming proton decay experiments, collider searches, and low energy experiments in search of flavor violations.   It is interesting to note that one also requires two copies of $15_H$s to correctly produce the matter-antimatter asymmetry of the universe. Moreover, implementing the type-III seesaw requires two copies of $24_F$ fermions, for which corners of the parameter space exist where a large gauge coupling unification scale can be obtained, making this scenario difficult to probe experimentally. For the latter two cases (i.e., \circled{2} and \circled{3}), we find that even with the implementation of the type-I seesaw to generate the neutrino masses, high scale unification can easily be achieved without requiring new physics states lower than $10^6$ GeV, making these scenarios difficult to probe experimentally. These findings are summarized in Table\ \ref{tab:max_proton_lifetime_models}.

This paper is organized in the following way. In Sec.~\ref{sec:5}, we propose the minimal model with a $5_F+\overline{5}_F$ VLFs and provide all the model details, including a phenomenological study of the model. In Sec.~\ref{sec:1015}, we explore the possibility of replacing the $5_F+\overline{5}_F$ VLFs with either a $10_F+\overline{10}_F$ or a $15_F+\overline{15}_F$. Finally, we conclude in Sec.~\ref{sec:con}.

\section{Case study: $5_F+\overline{5}_F$ VLFs}\label{sec:5}
As mentioned earlier, our goal is to build a viable minimal renormalizable model with representations no higher than  adjoints, i.e., $R \le 24$. In this section, we consider the case with a pair of $5_F+\overline{5}_F$ VLFs to resolve~\cite{Babu:2012pb} (see also~\cite{Dorsner:2014wva}) the bad mass relation. Within this setup,   if the type-I seesaw mechanism is employed for neutrino mass generation, the GUT scale comes out to be  $M_\mathrm{GUT}\lesssim 1.0\times 10^{14}$\ GeV, which is too low and is incompatible with current proton decay bounds. The minimal value of the GUT scale compatible with the current proton decay bound can be estimated as follows. From the superheavy gauge boson-mediated proton decay, the expected lifetime can be written as~\cite{Langacker:1980js}
\begin{align}
\tau_p\sim \frac{16\pi^2 M^4_X}{g^4_\mathrm{GUT}m^5_p},    
\end{align}
where $m_p$ and $M_X$ are the proton and the gauge boson masses, respectively, and $g_\mathrm{GUT}$ stands for the unified gauge coupling. Then, from the current proton decay bound of $\tau_p (p\to e^+\pi^0)> 2.4\times 10^{34}$ yrs, we obtain $M_X\sim M_\mathrm{GUT}\gtrsim 6\times 10^{15}$ GeV, where we have used $g_\mathrm{GUT}=0.6$.

For the type-II seesaw with one copy of $15_H$, we find the maximum possible GUT scale to be  $M_\mathrm{GUT}\lesssim 6.7\times 10^{14}$ GeV. This maximum value is also not compatible with the present experimental limits on proton decay. This is why, in the following, we study the scenario with two copies of $15_H$ Higgs fields, where the maximum unification scale we obtain is   $M_\mathrm{GUT}\lesssim 6.3\times 10^{15}$\ GeV (at two-loop order), making this scenario highly predictive as will be discussed in more detail later in the text.  Before presenting the details of this model, we point out that our study shows that if, on the other hand, the type-III seesaw mechanism is used, which in the absence of  $45_H$ Higgs requires at least two copies of fermionic $24_F$, assuming (nearly) mass degenerate weak triplets (which is required for resonant leptogenesis) a GUT scale of order $M_\mathrm{GUT}\lesssim 2\times 10^{15}$\ GeV (at two-loop order) can be obtained, which is a factor of 3 smaller than the expected lower limit mentioned above. If the assumption of degenerate weak triplet masses is dropped, the GUT scale can be as high as $M_\mathrm{GUT}\lesssim 9\times 10^{16}$\ GeV, making the model difficult to probe.

\subsection{Charged fermion masses}
As in the GG model, the GUT symmetry is spontaneously broken to the SM group via the vacuum expectation value (VEV) of the adjoint Higgs. Finally, the SM is broken at the electroweak (EW) scale when a Higgs in the fundamental representation acquires its VEV. These fields, under the SM group, decompose in the following way: 
\begin{align}
    &24_H=\phi_8(8,1,0)+\phi_1(1,3,0)+\phi_0(1,1,0)+\phi_3(3,2,-5/6)+\phi_{\overline{3}}(\overline{3},2,5/6),\\
    &5_H=H(1,2,1/2)+T(3,1,-1/3).
\end{align}
Moreover, the decomposition of the  VLF is shown below,  
\begin{align}
    &5_{F_4}=\overline{d^c_4}(3,1,-1/3)+\overline{L}_4(1,2,1/2),\\
    &\overline{5}_{F_4}=d^c_4(\overline{3},1,1/3)+L_4(1,2,-1/2).
\end{align}

With this set of fields, the complete Yukawa sector of the theory is~\cite{Babu:2012pb}
\begin{align}
-\mathcal{L}_Y=10_iY^{ij}_{10}10_j5_H+\overline 5_aY^{aj}_510_j5^*_H + \overline 5_a\left( \mu_a+\eta_a 24_H \right)5_4,    
\end{align}
where $i,j\in\{1,2,3\}$ and  $a,b\in\{1,2,3,4\}$ are the family indices.  Without loss of generality, one can choose a basis where  the upper $3\times 3$ block of $Y_5$ is real and diagonal in the family space, $Y_5=\mathrm{diag}\left(y_1\;y_2\;y_3\right)$. After the EW symmetry is broken, the mass terms for the  fermions can be written as 
\begin{align}
&-\mathcal{L}_Y=L^T M_E E^c+ D^TM_D D^c+ u^T M_U u^c, 
\end{align}
where the corresponding fields are defined in the following way:
\begin{align}
&L^T=(\ell_1,\ell_2,\ell_3,\ell_4), \;\; E^{cT}=(e^c_1,e^c_2,e^c_3,e^c_4),
\\
&D^T=(d_1,d_2,d_3,\overline d^c_4), \;\; D^{cT}=(d^c_1,d^c_2,d^c_3,d^c_4).
\end{align}
The $3\times 3$ mass matrix for the up-type quarks and $4\times 4$ matrices for the down-type quarks and charged leptons are given by 
\begin{align}
&M_U=4v_5(Y_{10}+Y^T_{10}),
\\
&M_D=\begin{pmatrix}
v_5 Y_5&0\\
\mu_i+2\eta_i v_{24}&|\mu_4+2\eta_4 v_{24}|
\end{pmatrix},
\\
&M_E=\begin{pmatrix}
v_5 Y_5&\mu_i-3\eta_i v_{24}\\
0&|\mu_4-3\eta_4 v_{24}|
\end{pmatrix}.
\end{align}
As expected, the up-type quark mass matrix is symmetric.   In the above equations  we have used the notation $\langle 5_H\rangle=v_5$ and  $\langle 24_H\rangle=  v_{24}(2,2,2,-3,-3)$, with $v_{24}=V_\mathrm{GUT}/\sqrt{15}$.  For later convenience, we further define $m_i=v_5y_i$, $M^D_a=\mu_a+2\eta_a v_{24}$, $M^E_a=\mu_a-3\eta_a v_{24}$ and $M_{L_4}=\sqrt{\sum_a|M_a^{E}|^2}$, $M_{d^c_4}=\sqrt{\sum_a|M_a^{D}|^2}$.

\subsection{Neutrino mass}
In our model, neutrino mass is generated by the type-II seesaw mechanism, for which we introduce scalars in the $15_H$ dimensional representation. As mentioned above, even though obtaining correct neutrino oscillation data requires one copy, too rapid proton decay rules out this scenario. Consequently, we introduce two copies of $15_H$. In the following analysis, we keep the index of this field implicit. A $15_H$ field decomposes in the following way:  
\begin{align}
    &15_{H}=\Delta_{1}(1,3,1)+\Delta_{3}(3,2,1/6)+\Delta_{6}(\overline{6},1,-2/3).
\end{align}
The weak triplet $\Delta_{1}(1,3,1)$ is responsible for generating neutrino masses via the type-II seesaw mechanism. Additionally, $15_H$ contains a scalar leptoquark $\Delta_{3}(3,2,1/6)$ commonly known as $\widetilde R_2$, and a scalar sextet $\Delta_{6}(\overline{6},1,-2/3)$. As we will see, this leptoquark (LQ) plays a crucial role in achieving unification at a high scale.

The additional terms in the Yukawa sector due to the presence of $15_H$ are, 
\begin{align}\label{eq:neutrino Yukawa couplings}
-\mathcal{L}_Y\supset \overline 5^a_F \overline 5^b_F 15_H Y^{ab}_{15} + 5^4_F5^4_F 15_H^\ast y^\prime,    
\end{align}
where $y^\prime$ is a number and $Y_{15}$ is a $4\times 4$ symmetric matrix. For the simplicity of the analysis, we assume that both $15_H$ fields share the same Yukawa coupling, and that sub-multiplets are degenerate in mass. The latter assumption is crucial in maximizing the GUT scale. Splitting their masses would only reduce the maximally allowed unification scale.  

The neutrino mass matrix then becomes a $5\times 5$ matrix, which in the   $(\nu_{L_a}, N^c_R)$ basis (where we adopt the notation that $\nu_{L_4}=N_L$ is the neutral  component of the extra left-handed fermion doublet $L_L(1,2,-1/2)=\left(N_L, E^-_L\right)^T$, and where $N_R$ is the corresponding neutral component in the right-handed doublet $L_R(1,2,-1/2)=\left(N_R, E^-_R\right)^T$)  takes the form 
\begin{align}
&M_N=\begin{pmatrix}
v_\Delta Y_{15}&M_a^E\\
M_b^E&v^\ast_\Delta y^\prime 
\end{pmatrix}_{5\times 5}=N^\ast M_N^\mathrm{diag}N^\dagger. \label{nuN}
\end{align}

Here we motivate the existence of two copies of $15_H$ representations. Although one copy of $15_H$ is enough to account for the neutrino oscillation data, it is not sufficient to produce the observed baryon asymmetry of the universe. In fact, one needs two such copies~\cite{Ma:1998dx,Hambye:2005tk,Lavignac:2015gpa}, as suggested by our proposed model.   Unlike the heavy Majorana neutrinos of standard leptogenesis~\cite{Fukugita:1986hr}, since the scalar
triplet $\Delta_1$ is not a self-conjugate state, one has both a triplet $\Delta_1$ and its anti-triplet $\overline \Delta_1$. Nevertheless, there is no CP asymmetry in $\Delta_1/\overline \Delta_1$ decays at the one-loop level. To have a non-zero CP asymmetry, one must have another state, e.g., another triplet, $\Delta_2$, with couplings to the lepton and Higgs doublets. Introducing this second copy of the triplet then yields one-loop processes that can contribute to sufficiently large CP asymmetries.

In the standard scenario of type-II seesaw leptogenesis,  it is typically assumed that $\Delta_2$ is much heavier than $\Delta_1$, and  the CP asymmetries in the decays of the triplets and anti-triplets are generated via their decays to SM leptons and an SM Higgs boson pair, $\Delta_1\to HH, \overline \ell\overline\ell$ and $\Delta_1\to \overline H\overline H, \ell\ell$. Without assuming extra
sources of CP-violation unrelated to neutrino masses, it is shown that a correct baryon asymmetry is obtained for a triplet mass of $m_{\Delta_1}\gtrsim 10^{10}$ GeV~\cite{Hambye:2005tk,Lavignac:2015gpa}, which is precisely what is predicted by our model from proton decay constraints (as shown later in the text). However, our scenario is more involved since additional $2\to 2$ scattering as well as decay channels  of the triplet are allowed and have more freedom compared to the vanilla scenario. Therefore,  we leave the study of leptogenesis for the future.

\subsection{Flavor violation}
It will be shown later that to maximize the GUT scale, the vectorlike quark (VLQ) needs to be at the GUT scale, while, on the contrary, the vectorlike doublet (VLD) needs to reside in the $1-100$ TeV range. Furthermore, the scalar LQ must live very close to the TeV scale to maximize the GUT scale and evade stringent proton decay constraints. This leads to interesting correlations between proton decay mediated by the GUT scale particles with the quark and lepton flavor violating processes mediated by the VLD and LQ residing at low scales. In this section, we compute their contributions to flavor violating processes.

First, we make a change of basis, 
\begin{align}
L^TM_E E^c\to \overline L_L \hat M_E E_R,\;\;\; \hat M_E=M^\ast_E    
\end{align}
and diagonalize this $4\times 4$ matrix as
\begin{align}
\hat M_E^{\textrm{diag}}=U^e_L\hat M_E U^{e\dagger}_R.  
\end{align}
Then the interactions of the charged lepton mass eigenstates (note the abuse of notation, i.e., flavor and mass eigenstates are denoted by the same symbols) with the $Z$ boson is given by
\begin{align}
&\mathcal{L}_Z\supset \bigg\{  
(g^Z_L)_{ab}\overline L_{L_a}\gamma^\mu L_{L_b} +
(g^Z_R)_{ab}\overline E_{R_a}\gamma^\mu E_{R_b} 
\bigg\}Z_\mu ,
\end{align}
where
\begin{align}
(g^Z_R)_{ab}= \frac{g}{c_W}s^2_W \delta_{ab}+
\frac{g}{2c_W}\bigg\{ U^e_R\;  \mathrm{diag}\left(0\;0\;0\;1\right) {U^e_R}^\dagger \bigg\}.
\end{align}
On the other hand, the corresponding left-handed interactions do not mediate flavor violation since $g^Z_L\propto 1_{4\times 4}$.

Similarly, we obtain the interactions with the  $W$ boson that lead to 
\begin{align}
&(g^W_L)_{\alpha a}= \frac{g}{\sqrt{2}} (P^\dagger)_{\alpha b} (R)_{ba},
\;\;\; 
(g^W_R)_{\alpha a}= \frac{g}{\sqrt{2}} (Q^\dagger)_\alpha (S)_a.
\end{align}
Here, we have defined the mixing matrices
\begin{align}
&P_{i\alpha}=\left(N_{3\times 5}\right)_{i\alpha},
\;
&&Q_{\alpha}=\left(N^\ast_{1\times 5}\right)_{5\alpha},
\;
&&R_{ia}=\left({U^e_L}^\dagger_{3\times 4}\right)_{ia},
\;
&&S_{a}=\left({U^e_R}^\dagger_{1\times 4}\right)_{4a},
\end{align}
and the index $\alpha$ takes the values  $\alpha\in\{1,...,5\}$. 

These interactions lead to cLFV in the form of $\ell\to \ell^\prime \gamma$, $\ell\to 3\ell^\prime$, and $\mu\to e$ conversion. The decay width of the $\ell\to \ell^\prime \gamma$ process is given by~\cite{Lavoura:2003xp}, 
\begin{align}
\Gamma(\ell_\alpha\to\ell_\beta \gamma)= 
\frac{\alpha_{em}m^3_{\ell_\alpha}}{1024e^2\pi^4}\bigg\{ 
\left| \sum_p A^p_{2L,\alpha\beta} \right|^2
+
\left| \sum_p A^p_{2R,\alpha\beta} \right|^2
\bigg\},
\end{align}
where $p$ runs over $W$ and $Z$ bosons. Processes of the type $\ell\to 3\ell^\prime$ are mediated by the $Z$ boson and have the following expressions~\cite{Kuno:1999jp,Porod:2014xia}:
\begin{align}
&\Gamma(\ell_\alpha\to 3\ell_\beta)= \frac{m^5_{\ell_\alpha}}{512\pi^3}
\bigg\{
\frac{2}{3}\left| T^{Z \ell\ell}_{RR,\alpha\beta\beta\beta}\right|^2
+
\frac{1}{3}\left| T^{Z \ell\ell}_{RL,\alpha\beta\beta\beta}\right|^2
\bigg\},
\\
&\Gamma(\ell^-_\alpha\to \ell^-_\beta\ell^-_\gamma\ell^+_\gamma)= \frac{m^5_{\ell_\alpha}}{512\pi^3}
\bigg\{
\frac{1}{3}\left| T^{Z \ell\ell}_{RR,\alpha\beta\gamma\gamma}\right|^2
+
\frac{1}{3}\left| T^{Z \ell\ell}_{RL,\alpha\beta\gamma\gamma}\right|^2
\bigg\},
\\
&\Gamma(\ell^-_\alpha\to \ell^+_\beta\ell^-_\gamma\ell^-_\gamma)= \frac{m^5_{\ell_\alpha}}{512\pi^3}
\bigg\{
\frac{2}{3}\left| T^{Z \ell\ell}_{RR,\alpha\gamma\beta\gamma}\right|^2
+
\frac{1}{3}\left| T^{Z \ell\ell}_{RL,\alpha\gamma\beta\gamma}\right|^2
\bigg\},
\end{align}
where we have defined
\begin{align}
&T^{Z \ell\ell}_{RR,jikl}=\frac{-1}{m^2_Z} \left(g^Z_R\right)_{ij}   \left(g^Z_R\right)_{lk},
\\&
T^{Z \ell\ell}_{RL,jikl}=\frac{-1}{m^2_Z} \left(g^Z_R\right)_{ij}   \left(g^Z_L\right)_{lk}.  
\end{align}
Furthermore, we find the following expressions for the relevant amplitudes: 
\begin{align}
&A^W_{2L,ji}=-2e\bigg\{
\left((g^W_R)^\ast_{\alpha i}(g^W_R)_{\alpha j}m_{e_i} 
+
(g^W_L)^\ast_{\alpha i}(g^W_L)_{\alpha j}m_{e_j}
\right)I^W_1
+
\left(3(g^W_L)^\ast_{\alpha i}(g^W_R)_{\alpha j}m_{\nu_\alpha}\right)I^W_2
\bigg\},
\end{align}
$A^W_{2R,ji}=A^W_{2L,ji}(L\leftrightarrow R)$, and $x=m^2_{\nu_\alpha}/m^2_W$. The $I^W_i$ functions are defined as 
\begin{align}
&I_1^W=\frac{6(1-3x)x^2\ln x+(x-1)[x(31x-26)+7]}{12m^2_W(x-1)^4},  
\\&I_2^W=\frac{2x^2\ln x+(4-3x)x-1}{2m^2_W(x-1)^3}. 
\end{align}

Similarly, for the $Z$ boson, 
\begin{align}
&A^Z_{2L,ji}=4e\bigg\{
-\left((g^Z_R)^\ast_{a i}(g^Z_R)_{a j}m_{e_i} 
\right)I_1^Z
+
2\left((g^Z_L)^\ast_{a i}(g^Z_R)_{a j}m_{e_a}\right)I_2^Z
\bigg\},
\\
&A^Z_{2R,ji}=4e\bigg\{
\left((g^Z_R)^\ast_{a i}(g^Z_R)_{a j}m_{e_i} 
\right)I_1^Z
+
2\left((g^Z_R)^\ast_{a i}(g^Z_L)_{a j}m_{e_a}\right)I_2^Z
\bigg\}
\end{align}
with $x=m^2_{e_a}/m^2_Z$, and
\begin{align}
&I_1^Z=\frac{-4+9x-5x^3+6x(2x-1)\ln x}{12m^2_Z(x-1)^4},   
\\&I_2^Z=\frac{-1+x^2-2x \ln x}{2m^2_Z(x-1)^3}.
\end{align}

Interactions of the $Z$ boson also lead to
$\mu\to e$ conversion that takes the form~\cite{Kitano:2002mt,Porod:2014xia}, 
\begin{align}
CR=\frac{m_\mu^5 \alpha_{em}^3}{16\pi^2}\left( \frac{Z^4_{\text{eff}}F^2_p}{Z\Gamma_\mathrm{cap}} \right)
\left| C^V_d\left( N G^{(d,n)}_V+Z G^{(d,p)}_V \right)
+
C^V_u\left( N G^{(u,n)}_V+Z G^{(u,p)}_V \right)\right|^2,\label{CR}
\end{align}
where, $C^V_q=T^{Z qq}_{RR, 21}+T^{Z qq}_{RL, 21}$, and we have defined, 
\begin{align}
&T^{Z dd}_{RR, ji}=\frac{-1}{m^2_Z} \left(g^Z_R\right)_{ij} g^{Z qq}_R, 
&T^{Z dd}_{RL, ji}=\frac{-1}{m^2_Z} \left(g^Z_R\right)_{ij} g^{Z qq}_L,  
\end{align}
and,
\begin{align}
&g^{Zuu}_L=- \frac{1}{6}\left(3g_2c_W-g_1s_W\right),
&g^{Zuu}_R=- \frac{2}{3}g_1s_W,
\\
&g^{Zdd}_L= \frac{1}{6}\left(3g_2c_W+g_1s_W\right), 
&g^{Zdd}_R=- \frac{1}{3}g_1s_W.  
\end{align}
In Eq.~\eqref{CR}, $Z$ and $N$ are the numbers of protons and neutrons in the nucleus, and $Z_{\text{eff}}$ is the effective atomic charge~\cite{chiang1993coherent}. $\Gamma_{\text{cap}}$ is the total muon capture rate, and $F_p$ represents the nuclear matrix element. The values of the relevant $G_V$ factors can be found in~\cite{Kuno:1999jp,Kosmas:2001mv}.

Finally, the scalar leptoquark contributes to both cLFV and semileptonic decays of kaons. We derive the following formulas for the relevant processes~\cite{Mandal:2019gff}, 
\begin{align}
&BR\left(K^0_L\to \mu^\pm e^\mp\right)= \frac{\tau_K f^2_K m^2_\mu m_{K^0}}{256\pi m^4_\mathrm{LQ}} 
\left(1-\frac{m^2_\mu}{m^2_K}\right)^2
\left| \hat Y_{21}\hat Y_{12}^\ast + \hat Y_{11}\hat Y_{22}^\ast \right|^2,
\end{align}
\begin{equation}
  \setlength{\arraycolsep}{0pt}
BR\left(K^+\to \pi^+\mu e\right)= \frac{\tau_K m^5_{K^+} |f(0)|^2 I_0}{12288\pi^3} 
            \left\{ \begin{array}{l l}
\left| \hat Y_{21}\hat Y_{12}^\ast \right|^2;\;\;\; K^+\to \pi^+\mu^+ e^-
    \\
\left| \hat Y_{22}\hat Y_{11}^\ast \right|^2;\;\;\; K^+\to \pi^+\mu^- e^+    
             \end{array} \right.,
\end{equation}
where we have defined $\hat Y=U^d_R Y_{15} {U^e_L}^\dagger$ and 
$I_0=0.178366$. Moreover, 
\begin{align}
CR=\frac{m_\mu^5 \alpha_{em}^3}{16\pi^2}\left( \frac{Z^4_{eff}F^2_p}{Z\Gamma_\mathrm{cap}} \right)
\left| C^V_d\left( N G^{(d,n)}_V+Z G^{(d,p)}_V \right) \right|^2,
\end{align}
with $C^V_d=T^{\phi dd}_{LR, 2111}$, and
\begin{align}
T^{\phi dd}_{LR, jikl}=\frac{-1}{2m^2_\mathrm{LQ}} \left(\hat Y^\ast\right)_{ik} \left(\hat Y\right)_{lj}.    
\end{align}

Current experimental bounds and future sensitives of these flavor violating processes are summarized in Table~\ref{tab:flavor_violation}.

\begin{table}[t!]
\centering
\begin{tabular}{|c|c|c|}\hline
Process  & Current bound & Future sensitivity  
\\\hline\hline
BR($\tau\to\mu\gamma)$ & $4.4\times 10^{-8}$\ \cite{BaBar:2009hkt} & $\sim 10^{-9}$\ \cite{Aushev:2010bq}
\\
BR($\tau\to e\gamma$) & $3.3\times 10^{-8}$\ \cite{BaBar:2009hkt} & $\sim 10^{-9}$\ \cite{Aushev:2010bq}
\\
BR($\mu\to e\gamma$) & $4.2\times 10^{-13}$\ \cite{MEG:2016leq} & $6\times 10^{-14}$\ \cite{Baldini:2013ke}
\\\hline
BR($\tau\to \mu\mu\mu)$ & $2.1\times 10^{-8}$\ \cite{Hayasaka:2010np} & $\sim 10^{-9}$\ \cite{Aushev:2010bq}
\\
BR($\tau\to e e e)$ & $2.7\times 10^{-8}$\ \cite{Hayasaka:2010np} & $\sim 10^{-9}$\ \cite{Aushev:2010bq}
\\
BR($\mu\to e e e)$ & $1.0\times 10^{-12}$\ \cite{SINDRUM:1987nra} & $\sim 10^{-16}$\ \cite{Blondel:2013ia}
\\
BR($\tau^-\to e^-\mu\mu$) & $2.7\times 10^{-8}$\ \cite{Hayasaka:2010np} & $\sim 10^{-9}$\ \cite{Aushev:2010bq}
\\
BR($\tau^-\to \mu^-ee$) & $1.8\times 10^{-8}$\ \cite{Hayasaka:2010np} & $\sim 10^{-9}$\ \cite{Aushev:2010bq}
\\
BR($\tau^-\to e^+\mu^-\mu^-$) & $1.7\times 10^{-8}$\ \cite{Hayasaka:2010np} & $\sim 10^{-9}$\ \cite{Aushev:2010bq}
\\
BR($\tau^-\to \mu^+e^-e^-$) & $1.5\times 10^{-8}$\ \cite{Hayasaka:2010np} & $\sim 10^{-9}$\ \cite{Aushev:2010bq}
\\
\hline
CR($\mu \textrm{Au}\to e \textrm{Au})$ & $7\times 10^{-13}$\ \cite{SINDRUMII:2006dvw} & $-$
\\
CR($\mu \textrm{Ti}\to e \textrm{Ti})$ & $4.3\times 10^{-12}$\ \cite{SINDRUMII:1993gxf} & $\sim 10^{-18}$\ \cite{unPUB}
\\
CR($\mu \textrm{Al}\to e \textrm{Al})$ & $-$ & $10^{-15}-10^{-18}$\ \cite{Pezzullo:2017iqq}
\\
\hline
BR($K_L^0\to\mu^\pm e^\mp$) & $4.7\times 10^{-12}$\ \cite{BNL:1998apv} & $\sim 10^{-12}$\ \cite{Goudzovski:2022vbt}
\\
BR($K_L^0\to\pi^0\mu^+ e^-$) & $7.6\times 10^{-11}$\ \cite{KTeV:2007cvy} & $\sim 10^{-12}$\ \cite{Goudzovski:2022vbt}
\\
BR($K^+\to\pi^+\mu^+ e^-$) & $1.3\times 10^{-11}$\ \cite{Sher:2005sp} & $\sim 10^{-12}$\ \cite{Goudzovski:2022vbt}
\\
BR($K^+\to\pi^0\mu^- e^+$) & $5.2\times 10^{-10}$\ \cite{Appel:2000tc} & $\sim 10^{-12}$\ \cite{Goudzovski:2022vbt}
\\
\hline
\end{tabular}
\caption{Current experimental constraints and future sensitivities for various lepton violating processes and semileptonic decays of kaons, all at the 90\% confidence level.}\label{tab:flavor_violation}
\end{table}

\subsection{Gauge coupling unification}\label{sec:gauge coupling unification}
In order to perform the gauge coupling unification analysis we compute the renormalization group (RG) running of the SM gauge couplings at two-loop. The corresponding beta functions (with $i\in\{1,2,3\}$) read 
\begin{align}
\label{eq:2loop_gauge_running}
    \mu\frac{d\alpha_i^{-1}}{d\mu}=
    &-\frac{1}{2\pi}\left(a_i^{\textrm{SM}}+\sum_J a_i^J\theta(\mu,M_J)\right)
    \nonumber\\
    &-\frac{1}{8\pi^2}\left(\sum_j\Big(b_{ij}^{\textrm{SM}}+\sum_J b_{ij}^J\theta(\mu,M_J)\Big)\alpha_j^{-1}+\beta_i^Y\right),
\end{align}
with $a_i^\textrm{SM}$ ($b_{ij}^\textrm{SM}$) being the SM one-loop (two-loop) gauge coefficients, whereas $a_i^J$ ($b_{ij}^J$) are the one-loop (two-loop) gauge coefficients of the multiplets $J$ with masses $M_J$, such that $M_Z\leq M_J\leq M_\textrm{GUT}$. The gauge coefficients $a_i^J$ and $b_{ij}^J$ can be found in Appendix\ \ref{app:RGE}. Moreover, $\theta(\mu,m)=1$ if $\mu>m$ and $\theta(\mu,m)=0$ if $\mu\leq m$, denotes the step function. The Yukawa contributions $\beta_i^Y$ are neglected for beyond the SM Yukawa couplings. 

\begin{figure}
    \centering
    \includegraphics[width=11cm]{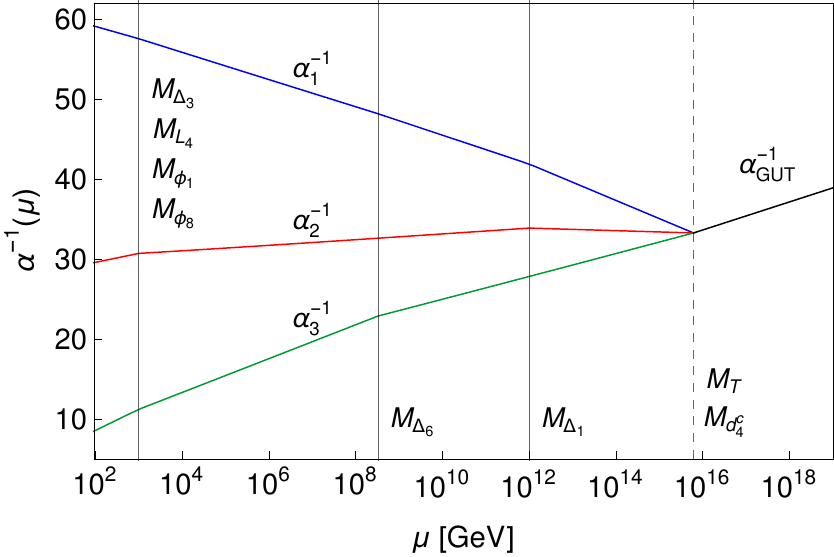}
    \caption{Example for gauge coupling unification at two-loop and a GUT scale of $6.15\times 10^{15}$\ GeV. The particle masses are chosen as $M_{\Delta_1}=1.00\times 10^{12}$\ GeV and $M_{\Delta_6}=3.28\times 10^8$\ GeV.}
    \label{fig:gauge coupling unification}
\end{figure}

To achieve gauge coupling unification we freely vary all intermediate scale particle masses, i.e.\ the masses of the fields $\phi_8$, $\phi_1$, $T$, $L_4+\overline{L}_4$, $d^c_4+\overline{d^c_4}$, $\Delta_1$, $\Delta_3$, and $\Delta_6$. For the mass of the scalar color triplet $T$ we take a lower bound of $3\times 10^{11}$\ GeV to sufficiently suppress LQ mediated nucleon decay. The masses of all of the other fields are varied between the TeV and the GUT scale, while ensuring that all neutrino Yukawa couplings in Eq.\ \eqref{eq:neutrino Yukawa couplings} can be chosen perturbatively. We run the SM gauge couplings from the GUT scale down to the $Z$ scale, where we compute a $\chi^2$ function comparing the obtained values with the experimental low-scale values $g_1=0.461425$, $g_2=0.65184$, and $g_3=1.2143$\ \cite{Antusch:2013jca}, where we have used the relation $g_i=\sqrt{4\pi\alpha_i}$. Gauge coupling unification can, for example, be achieved if the intermediate scale particle masses are chosen as $M_{\phi_8}=1.00$\ TeV, $M_{\phi_1}=1.00$\ TeV, $M_{T}=M_\textrm{GUT}$, $M_{L_4}=1.00\ TeV$, $M_{d_4^c}=M_\textrm{GUT}$, $M_{\Delta_1}=1.00\times 10^{12}$\ GeV, $M_{\Delta_3}=1.00$\ TeV, and $M_{\Delta_6}=3.28\times 10^8$\ GeV. For this scenario we find a GUT scale of $M_\textrm{GUT}=6.15\times 10^{15}$\ GeV which is large enough to evade the current proton decay bounds. The corresponding gauge coupling unification plot is presented in Figure\ \ref{fig:gauge coupling unification}. Note that if the RG evolution is computed at one-loop the GUT scale cannot be larger than $3\times 10^{15}$\ GeV, which is roughly a factor of 2 too small to evade the proton decay constraints. This means that in order to show that our model is indeed viable a more accurate two-loop computation is required.

\subsection{Proton decay}
Here, we collect relevant formulas for computing proton decay rates. Decay widths for proton decay channels into charged anti-leptons and anti-neutrinos are given by (cf.\ \cite{Claudson:1981gh, Kuramashi:2000hw} for the remaining decay channels)
\begin{align}\label{eq:Gamma p->pi e}
\Gamma (p \to \pi^0 e^+_\beta)&= 
\frac{ m_p\pi}{2}\left(1-\frac{m_\pi^2}{m_p^2}\right)^2\frac{\alpha_\textrm{GUT}^2}{M_\textrm{GUT}^4}A_L^2\\\nonumber
    &\times\left(A_{SL}^2|c(e_\alpha^c,d)\langle\pi^0|(ud)_Lu_L|p\rangle|^2+A_{SR}^2|c(e_\alpha,d^c)\langle\pi^0|(ud)_Ru_L|p\rangle|^2\right),
    \\ \label{eq:Gamma p->K nu}
\Gamma (p \to K^+ \bar{\nu})&= 
\frac{ m_p\pi}{2}\left(1-\frac{m_{K^+}^2}{m_p^2}\right)^2\frac{\alpha_\textrm{GUT}^2}{M_\textrm{GUT}^4}A_L^2A_{SR}^2\\\nonumber 
&  \times 
\left(   \sum_i  \left|  
c(\nu_i,d,s^c)  \matrixel{K^+}{(us)_R d_L}{p}
+ c(\nu_i,s,d^c)  \matrixel{K^+}{(ud)_R s_L}{p} \right|^2  \right)\;\nonumber,
\\ \label{eq:Gamma p->eta e}
    \Gamma (p \to \eta^0 e^+_\beta)&= 
\frac{ m_p\pi}{2}\left(1-\frac{m_\eta^2}{m_p^2}\right)^2\frac{\alpha_\textrm{GUT}^2}{M_\textrm{GUT}^4}A_L^2
\\\nonumber
    &\times\left(A_{SL}^2|c(e_\alpha^c,d)\langle\eta^0|(ud)_Lu_L|p\rangle|^2+A_{SR}^2|c(e_\alpha,d^c)\langle\eta^0|(ud)_Ru_L|p\rangle|^2\right),
\end{align}
where $A_L=1.2$\ \cite{Nihei:1994tx} and $A_{SL(R)}$ denote the leading log dimension six operator renormalization. The latter is given by\footnote{A different factor, namely $\exp[\gamma_{L(R)i}\alpha(M_{J+1})]/(2\pi)$, is used instead if the one-loop gauge coefficient vanishes in a certain interval.}\ \cite{Wilczek:1979hc,Buras:1977yy,Ellis:1979hy}
\begin{align}
    &A_{SL(R)}=\prod_{i=1,2,3}\prod_{I}^{M_Z\leq M_I\leq M_\textrm{GUT}}\left(\frac{\alpha_i(M_{I+1})}{\alpha_i(M_I)}\right)^{\frac{\gamma_{L(R)i}}{b_i^\textrm{SM}+\sum_J^{M_Z\leq M_J\leq M_\textrm{GUT}}b_i^J}},\nonumber
    \\
    &\text{with\qquad $\gamma_{L(R)i}=\left(23(11)/20,9/4,2\right)$.}
\end{align}
Moreover, $m_p=939.3$\ MeV, $m_\pi=139.6$\ MeV, $m_{K^+}=493.7$\ MeV, and $m_\eta=547.9$\ MeV are the proton, pion, kaon, and eta meson mass, respectively. 
Taking into account the fact that in our model the up-type Yukawa matrix is symmetric, the c-coefficients read\footnote{Note the fact that the c-coefficients are modified in our model compared to their typical form due to the additional mixing with the VLFs.}\ \cite{DeRujula:1980qc,FileviezPerez:2004hn,Nath:2006ut}
\begin{align}
&c(e_\alpha^c,d_\beta)=(E_R^\ast)_{i\alpha} (D_L^\ast)_{i\beta}+(E_R^\ast)_{i\alpha} (U_L^\ast)_{i1}(U_L)_{i1} (D_L^\ast)_{i\beta}+(E_R^\ast)_{4\alpha}(D_L^\ast)_{4\beta},\\
    &c(e_\alpha,d^c_\beta)=(E_L^\ast)_{a\alpha} (D_R^\ast)_{a\beta},\\
    &c(\nu_l,d_\alpha,d^c_\beta)=(U_L)_{i1} (D_L^\ast)_{i\alpha}\left[(D_R^\ast)_{a \beta} (N)_{a l}+(D_R^\ast)_{4\beta}N_{5l}\right],
\end{align}
where we implicitly sum over the indices $i\in\{1,2,3\}$ and $a\in\{1,2,3,4\}$. The unitary matrices $U$, $D_L$, $D_R$, $E_L$, $E_R$, and $N$ are defined such that they diagonalize the corresponding fermion mass matrices
\begin{align}
    &M_U=U M_U^\textrm{diag} U^T,
    && M_D=D_LM_D^\textrm{diag}D_R^\dagger ,\nonumber
    \\
    &M_E=E_LM_E^\textrm{diag}E_R^\dagger,
    &&M_N=N^\ast M_N^\textrm{diag}N^\dagger.
\end{align}
Finally, the matrix elements are given by\ \cite{Aoki:2017puj,Yoo:2021gql}
\begin{align}
&\matrixel{\pi^0}{(ud)_Lu_L}{p}=+0.134(5)(16) \text{ GeV}^2,&& \matrixel{\pi^0}{(ud)_Ru_L}{p}=-0.131(4)(13) \text{ GeV}^2,\nonumber \\
&\matrixel{K^+}{(ud)_R s_L}{p} =-0.134(4)(14) \text{ GeV}^2,&& \matrixel{K^+}{(us)_R d_L}{p}=-0.049(2)(5) \text{ GeV}^2,\nonumber
\\
&\matrixel{\eta^0}{(ud)_Lu_L}{p}=+0.134(5)(16) \text{ GeV}^2,&& \matrixel{\eta^0}{(ud)_Ru_L}{p}=-0.131(4)(13) \text{ GeV}^2.
\end{align}
In Table\ \ref{tab:nucleon_decay} we show the present experimental bounds together with the future sensitivities for partial proton lifetimes for various decay channels.

\begin{table}[t!]
\centering
\begin{tabular}{|c|c|c|}\hline
Decay channel  & Current bound $\tau_p$ [yrs] & Future sensitivity $\tau_p$ [yrs] 
\\\hline\hline
$p\rightarrow \pi^0\,e^+$  &  $2.4\times 10^{34}$ \cite{Super-Kamiokande:2020wjk} &  $7.8\times 10^{34}$ \cite{Hyper-Kamiokande:2018ofw}   \\\hline
$p\rightarrow \pi^0\,\mu^+$  &  $1.6\times 10^{34}$ \cite{Super-Kamiokande:2020wjk} &  $7.7\times 10^{34}$ \cite{Hyper-Kamiokande:2018ofw}    \\\hline
$p\rightarrow \eta^0\,e^+$  &  $1.0\times 10^{34}$ \cite{Super-Kamiokande:2017gev} &  $4.3\times 10^{34}$ \cite{Hyper-Kamiokande:2018ofw}  \\\hline
$p\rightarrow \eta^0\,\mu^+$  &  $4.7\times 10^{33}$ \cite{Super-Kamiokande:2017gev} &  $4.9\times 10^{34}$ \cite{Hyper-Kamiokande:2018ofw}   \\\hline
$p\rightarrow K^0\,e^+$  &  $1.1\times 10^{33}$ \cite{Brock:2012ogj} &  -  \\ \hline
$p\rightarrow K^0\,\mu^+$  &  $3.6\times 10^{33}$ \cite{Super-Kamiokande:2022egr} &  -  \\\hline
$p\rightarrow \pi^+\,\overline{\nu}$  &  $3.9\times 10^{32}$ \cite{Super-Kamiokande:2013rwg} &  -  \\\hline
$p\rightarrow K^+\,\overline{\nu}$  &  $6.6\times 10^{33}$ \cite{Takhistov:2016eqm} &  $3.2\times 10^{34}$ \cite{Hyper-Kamiokande:2018ofw} \\
\hline
\end{tabular}
\caption{Table of current experimental bounds and future sensitivities (for 10 years of runtime) for different proton decay channels. For a recent review on the subject, see Ref.~\cite{Dev:2022jbf} }\label{tab:nucleon_decay}
\end{table}

\subsection{Numerical analysis}
This section is devoted to a step by step description of our numerical procedure. We start by parametrizing the fermion mass matrices. Already at this step we compute the unitary matrices that diagonalize the fermion mass matrices. These unitary matrices are used later on for the computation of proton decay and flavor violation predictions as well as for the fit of the Cabibbo-Kobayashi-Maskawa (CKM) and  Pontecorvo-Maki-Nakagawa-Sakata (PMNS) matrices. Moreover, the singular values of the mass matrices are used for the fermion mass fit.

To parametrize the down-type and charged lepton mass matrices we use the mass parameters $m_i$, the VLF mass $M_{d_4^c}$ ($M_{L_4}$), as well as the angles $\theta_i^{D,E}$ that are defined in Appendix\ \ref{app:Block diagonalization of fermion mass matrices}. We then reconstruct the mass parameters $M_a^{D,E}$ using Eq.\ \eqref{eq:relation theta_i^{D,E} - M_i^{D,E}}. It turns out that in order to allow for gauge coupling unification the VLQ stemming from $5_{F_4}+\overline{5}_{F_4}$ has to reside close to the GUT scale, i.e., its mass is 15 orders of magnitude above the bottom quark mass. Because of this, we can safely use the method described in Appendix\ \ref{app:Block diagonalization of fermion mass matrices} to block diagonalize $M_D$. Afterwards, we diagonalize the remaining upper $3\times 3$ block numerically. The VLL mass has to be close to the TeV scale to successfully achieve gauge coupling unification. That means that for the charged lepton mass matrix we cannot safely use the block diagonalization described in Appendix\ \ref{app:Block diagonalization of fermion mass matrices}, since it is only correct up to corrections of order $m_3/M_{L_4}$. We therefore diagonalize $M_E$ using a numerical method.

Utilizing the fact that the up-type mass matrix is symmetric in our model, we decompose it by a Takagi decomposition, 
\begin{align}\label{eq:parametrization of MU}
    M_U=U\text{diag}(m_u,m_c,m_t)U^T,
\end{align}
where $U$ is a unitary matrix. For the three up-type quark masses appearing in Eq.\ \eqref{eq:parametrization of MU} we directly insert their experimental central GUT scale values which we take from\ \cite{Babu:2016bmy}. Since the down-type quark mass matrix $M_D$ can be block diagonalized with very high accuracy using only a right rotation matrix, the left rotation matrix turns out to be non-trivial only in its upper $3\times 3$ block, i.e.,
\begin{equation}
    D_L=\begin{pmatrix}
        D_L^{3\times 3} & 0 \\ 
        0 & 1
    \end{pmatrix}.
\end{equation}
Therefore, the CKM matrix is approximately unitary and we can thus parametrize the unitary matrix $U$ as
\begin{align}
    U=D_L^{3\times 3}\textrm{diag}(e^{i\beta_1^u},e^{i\beta_2^u},1)V_\textrm{CKM}^\textrm{exp}\textrm{diag}(e^{i\eta_1^u},e^{i\eta_2^u},e^{i\eta_3^u}),
\end{align}
where $V_\textrm{CKM}^\textrm{exp}$ is the CKM matrix and where $\eta_i^u$ are unphysical parameters that can safely be set to zero, while the so-called GUT phases $\beta_1^u$ and $\beta_2^u$ are free phases that affect the proton decay predictions. We directly insert the experimental central GUT scale values into $V_\textrm{CKM}^\textrm{exp}$.

We block diagonalize the neutrino mass matrix using the method described in Appendix\ \ref{app:Block diagonalization of fermion mass matrices}, since $M_{L_4}$ is expected to reside at the TeV scale, while the other entries in $M_N$ are of the order of eV, i.e. the corrections to the approximate block diagonalization are of order $10^{-12}$. After the block diagonalization we decompose the symmetric upper $3\times 3$ block utilizing a Takagi decomposition
\begin{align}
    M_N^{3\times 3}={U_N^{3\times 3}}^\ast (m_{\nu_1},m_{\nu_2},m_{\nu_3}){U_N^{3\times 3}}^\dagger,
\end{align}
where $U_N^{3\times 3}$ is a unitary matrix. We take $m_{\nu_1}$ as a free parameter and use experimental central values of the two mass squared differences from NuFIT 5.2\ \cite{Esteban:2020cvm, NUFIT} to directly obtain $m_{\nu_2}$ and $m_{\nu_3}$. 

Since the rightmost columns in both mass matrices $M_E$ and $M_N$ depend on the same parameters, when computing the PMNS matrix $V_\textrm{PMNS}$ (which is a $4\times 5$ matrix)
\begin{align}\label{eq:VPMNS_4x5}
    (V_\textrm{PMNS})_{al}=(E_L^\ast)_{ba} N_{bl}=(E_L^\ast)_{ba}(P_L^N)_{bm}(V_N)_{mn}
    \begin{pmatrix}
        U_N^{3\times 3} & 0 \\
        0 & 1^{2\times 2}
    \end{pmatrix}_{nl}.
\end{align}
Therefore, defining
\begin{align}\label{eq:VEN_3x3}
    (V_{EN}^{3\times 3})_{ij}=(E_L^\ast)_{bi}(P_L^N)_{bm}(V_N)_{mj},
\end{align}
we parametrize $U_N^{3\times 3}$ as
\begin{align}\label{eq:UN_3x3}
    (U_N^{3\times 3})_{ij}=({V_{EN}^{3\times 3}}^{-1})_{ik} \left[\textrm{diag}(e^{i\eta_1^\nu},e^{i\eta_2^\nu},e^{i\eta_3^\nu})V_\textrm{PMNS}^\textrm{exp}\textrm{diag}(e^{i\beta_1^\nu},e^{i\beta_2^\nu},1)\right]_{kj},
\end{align}
where we plug the experimental central values of the PMNS parameters from NuFIT 5.2\ \cite{Esteban:2020cvm, NUFIT} into $V_\textrm{PMNS}^\textrm{exp}$. The phases $e^{i\beta_1^\nu}$, $e^{i\beta_2^\nu}$ denote the Majorana phases, while the phases $e^{i\eta_1^\nu}$, $e^{i\eta_2^\nu}$, $e^{i\eta_3^\nu}$ are unphysical and thus set to zero. In Eqs.\ \eqref{eq:VPMNS_4x5}-\eqref{eq:UN_3x3} the indices $i,j,k$ run from 1 to 3, the indices $a,b,c$ run from 1 to 4 and the indices $l,m,n$ run from 1 to 5.

In summary, to parametrize the fermion mass matrices we use the three mass parameters $m_i$, the six angles $\theta_i^{D,E}$, the nine phases in $M_D$ and $M_E$, the four phases $\beta_1^u$, $\beta_2^u$, $\beta_1^\nu$, $\beta_2^\nu$, and the light neutrino mass $m_{\nu_1}$. Additional parameters of the model are the GUT scale $M_\textrm{GUT}$, the unified gauge coupling $\alpha_\textrm{GUT}$, the masses of the intermediate scale fields $\phi_8$, $\phi_1$, $T$, $L_4+\overline{L}_4$, $d^c_4+\overline{d^c_4}$, $\Delta_1$, $\Delta_3$, $\Delta_6$, and the triplet Higgs VEV $v_\Delta$. Taking proton decay and flavor violation constraints into account, these parameters are fitted to the down-type quark and charged lepton masses $m_d$, $m_s$, $m_b$, $m_e$, $m_\mu$, $m_b$, and to the SM gauge couplings $g_1$, $g_2$, $g_3$, while also ensuring perturbativity of all Yukawa couplings. The remaining experimental values, namely the up-type quark masses, the CKM and PMNS parameters, as well as the neutrino mass squared differences are automatically accounted for. Note that in order to find a benchmark point with a good fit, not all input parameters need to be varied. In particular, fixed values can be assigned to all phases as well as to the light neutrino mass $m_{\nu_1}$. 

In the fitting procedure, we compute the two-loop running of the gauge couplings from the GUT scale down to the Z scale as discussed in Section\ \ref{sec:gauge coupling unification}. The gauge couplings are then fitted to their low energy values that we take from\ \cite{Antusch:2013jca}. The fermion masses and mixings, on the other hand, are for simplicity directly fitted at the GUT scale to their corresponding high energy values, which were provided in\ \cite{Babu:2016bmy}. For the experimental bounds on flavor violation and nucleon decay we use the values given in Tables\ \ref{tab:flavor_violation} and\ \ref{tab:nucleon_decay}, respectively. We then compute the $\chi^2$-function summing over the individual pulls $\chi_i^2$ for all observables $i$.
For the PMNS observables $\theta_{23}^\textrm{PMNS}$ and $\delta^\textrm{PMNS}$ we use the exact $\chi_i^2$ provided by NuFIT 5.2\ \cite{Esteban:2020cvm, NUFIT}. For all other observables $i$, we compute the pull $\chi_i^2$ via
\begin{align}
    \chi_i^2=\left(\frac{p_i-e_i}{\sigma_i}\right)^2,
\end{align}
where $p_i$ denotes the theoretical prediction, $e_i$ is the experimental central value, and $\sigma_i$ is the standard deviation. We obtain a viable benchmark point of the model minimizing the $\chi^2$-function using a differential evolution algorithm. Afterwards, we determine the posterior density of the observables of our model by applying an adaptive Metropolis-Hastings algorithm to perform an Markov chain Monte Carlo (MCMC) analysis. We start this MCMC analysis from the benchmark point and compute $6\times 10^6$ data points using flat prior probability distributions.

\subsection{Results}\label{sec:results}
In this section we present and discuss the results of our numerical analysis. We are mainly interested in the predictions for the rates of proton decay channels and their connection to the masses of the added scalar multiplets. Moreover, various predictions for flavor violating processes give rise to additional possibilities to test our model. 

\begin{figure}[t]
    \centering
    \includegraphics[width=14cm]{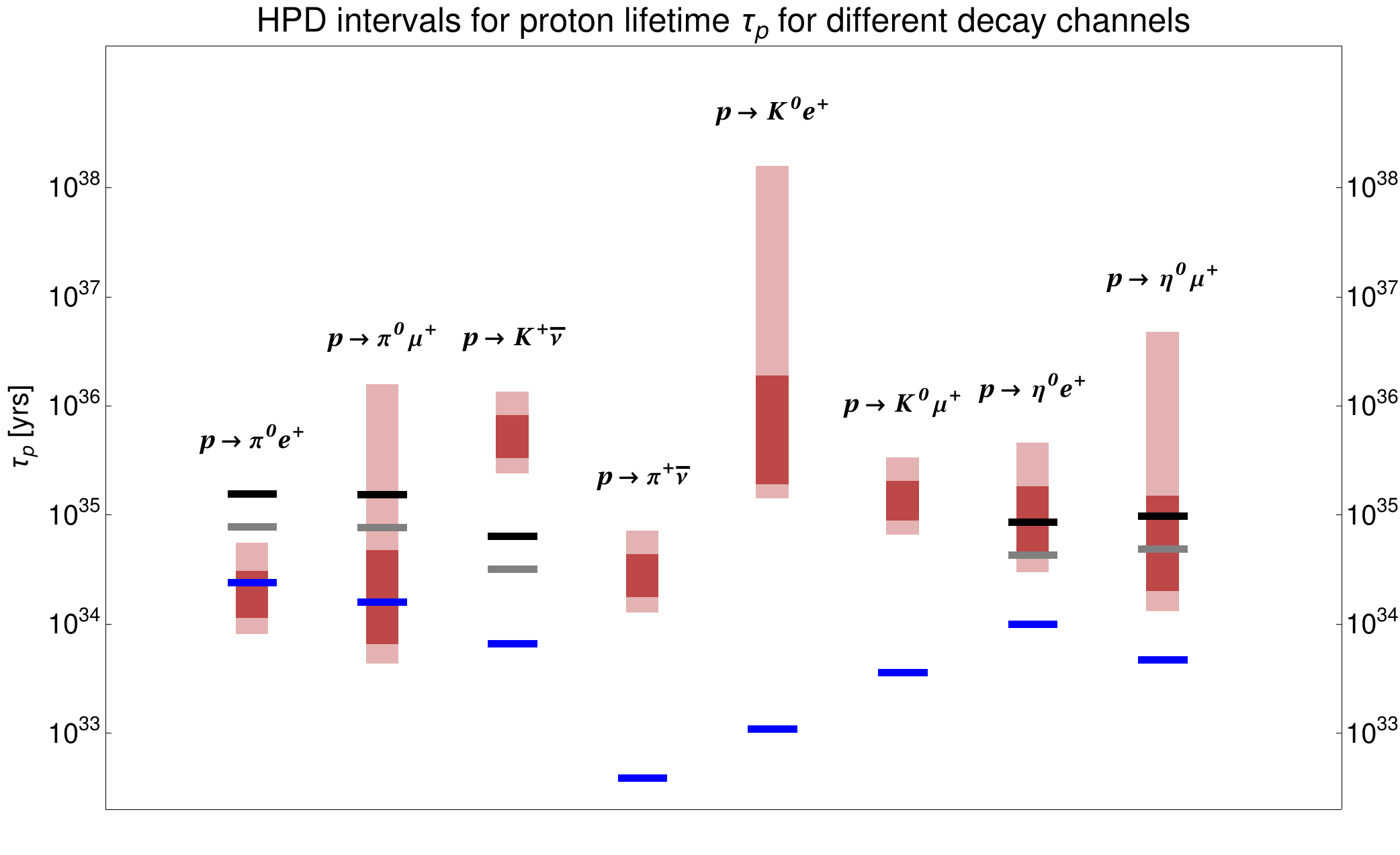}
    \caption{1$\sigma$ (dark) and 2$\sigma$ (light) HPD intervals of partial proton lifetimes for various decay channels. The current experimental bounds at the 90\% confidence level are indicated by blue line segments. Moreover, the future sensitivities  at the 90\% confidence level of Hyper-Kamiokande after a runtime of 10 years (20 years) are represented by gray (black) line segments.}
    \label{fig:HPD proton decay}
\end{figure}

We have presented in Section\ \ref{sec:gauge coupling unification} a possibility for achieving gauge coupling unification. If additionally, the mass matrices $M_D$ and $M_E$ are chosen as
\begin{align}
    &M_D=\begin{pmatrix}
        1.80\times 10^{-3} & 0 & 0 & 0 \\
        0 & 1.34\times 10^{-1} & 0 & 0 \\
        0 & 0 & 3.32 & 0 \\
        2.31\times 10^{14} & 1.84\times 10^{15} & 6.00\times 10^{15} & 1.90\times 10^{14}
    \end{pmatrix}\;\textrm{GeV},\\
    &M_E=\begin{pmatrix}
        1.80\times 10^{-3} & 0 & 0 & 5.77\times 10^2 \\
        0 & 1.34\times 10^{-1} & 0 & 5.40\times 10^2 \\
        0 & 0 & 3.32 & 1.36\times 10^3 \\
       0 & 0 & 0 & 1.56\times 10^2
    \end{pmatrix}\; \textrm{GeV},
\end{align}
then the down type and charged lepton masses can be fitted. This defines a viable benchmark point (with $\chi^2<10^{-2}$). We start our Markov chains from this benchmark point to approximate the posterior density. From the obtained points we compute the highest posterior density (HPD) intervals of partial proton lifetimes of various decay channels. Our findings are presented in Fig.\ \ref{fig:HPD proton decay}. The dark (light) rectangles represent the 1$\sigma$ (2$\sigma$) HPD intervals of partial proton lifetimes. The blue line segments represent the current experimental bounds, whereas the future sensitivities for a runtime of 10 years (20 years) are indicated by gray (black) line segments. Interestingly, Hyper-Kamiokande will be able to test four different proton decay channels; after a runtime of 10 years, it will already test the full $2\sigma$ HPD region of the decay channel $p\rightarrow \pi^0e^+$ as well as the full $1\sigma$ HPD interval of the decay channel $p\rightarrow \pi^0\mu^+$. Moreover, Hyper-Kamiokande will test part of the $1\sigma$ region of the two decay channels $p\rightarrow \eta^0 e^+$ and $p\rightarrow \eta^0 \mu^+$. 

\begin{figure}
    \centering
    \includegraphics[width=\textwidth/2-3mm]{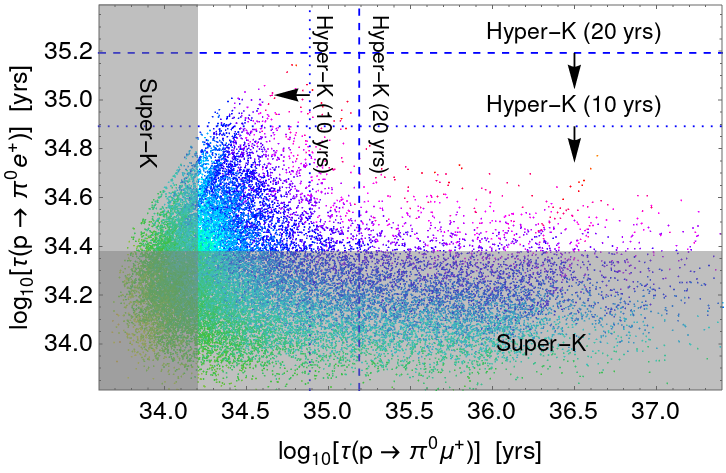}\hspace{3mm}
    \includegraphics[width=\textwidth/2-3mm]{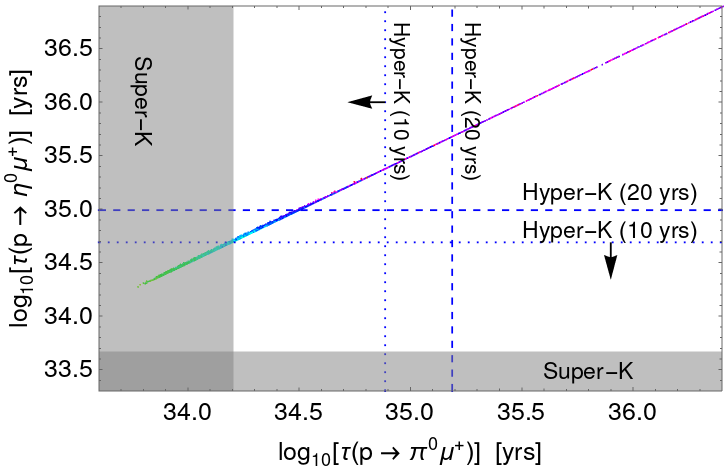}\\
    \includegraphics[width=\textwidth/3]{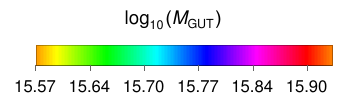}
    \caption{Left panel: Relationship between partial lifetimes of the proton decay channels $p\rightarrow \pi^0\mu^+$ and $p\rightarrow \pi^0e^+$. The two channels are inversely correlated for a fixed GUT scale. Right panel: Relationship between partial lifetimes of the proton decay channels $p\rightarrow \pi^0\mu^+$ and $p\rightarrow \eta^0e^+$. The two channels are highly correlated. Their ratio is predicted to lie within 3.09 and 3.47 at 2$\sigma$.}
    \label{fig:proton decay ratios}\vspace{15mm}
    \includegraphics[width=\textwidth/2-3mm]{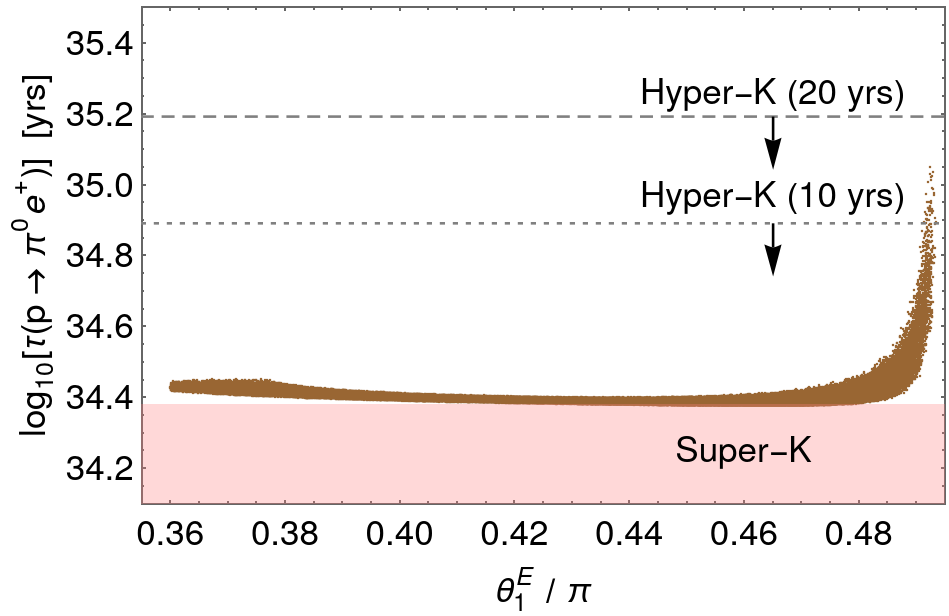}\hspace{3mm}
    \includegraphics[width=\textwidth/2-3mm]{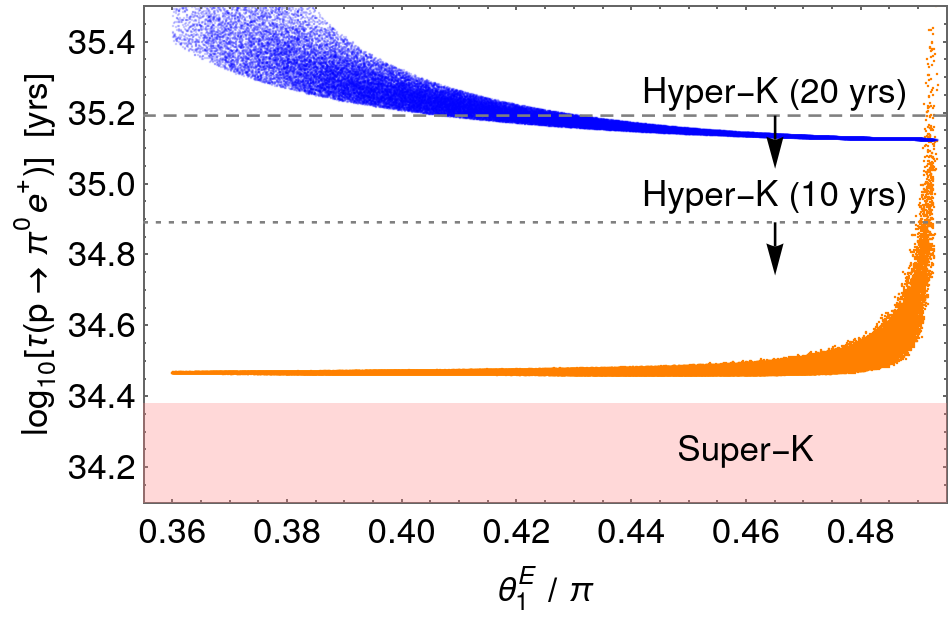}
    \caption{Dependence of the partial proton lifetime $\tau(p\rightarrow\pi^0e^+)$ on the model parameter $\theta_1^E$ for a benchmark scenario with $M_\textrm{GUT}=6.28\times 10^{15}$\ GeV. Left panel:  partial lifetime $\tau(p\rightarrow\pi^0e^+)$. Right panel: Individual contributions to the partial lifetime (cf.\ Eq.\ \eqref{eq:Gamma p->pi e}). The orange colored points indicate the contribution proportional to $c(e^c,d)$, while the blue colored points represent the contribution proportional to $c(e,d^c)$. See main text for details. }
    \label{fig:proton decay vs theta}
\end{figure}

Considering the ratios of two different decay channels, we find another interesting result which we present in Fig.\ \ref{fig:proton decay ratios}. In the left panel we show that for a given GUT scale the decay channels $p\rightarrow \pi^0 e^+$ and $p\rightarrow \pi^0 \mu^+$ are inversely correlated. This is particularly interesting since it tells us that, if the proton decay in the decay channel $p\rightarrow \pi^0 e^+$ is not observed after a 10 year runtime of Hyper-Kamiokande, we should definitely see proton decay in the decay channel $p\rightarrow \pi^0 \mu^+$ after a 20 year runtime of Hyper-Kamiokande. In the right panel we see that the two decay channels $p\rightarrow \eta^0 \mu^+$ and $p\rightarrow \pi^0 \mu^+$ are highly correlated. Our analysis finds their ratio at $2\sigma$ to lie within 3.09 and 3.47, which is another possibility to test our model. 

\begin{figure}[p]
    \centering
    \includegraphics[width=14cm]{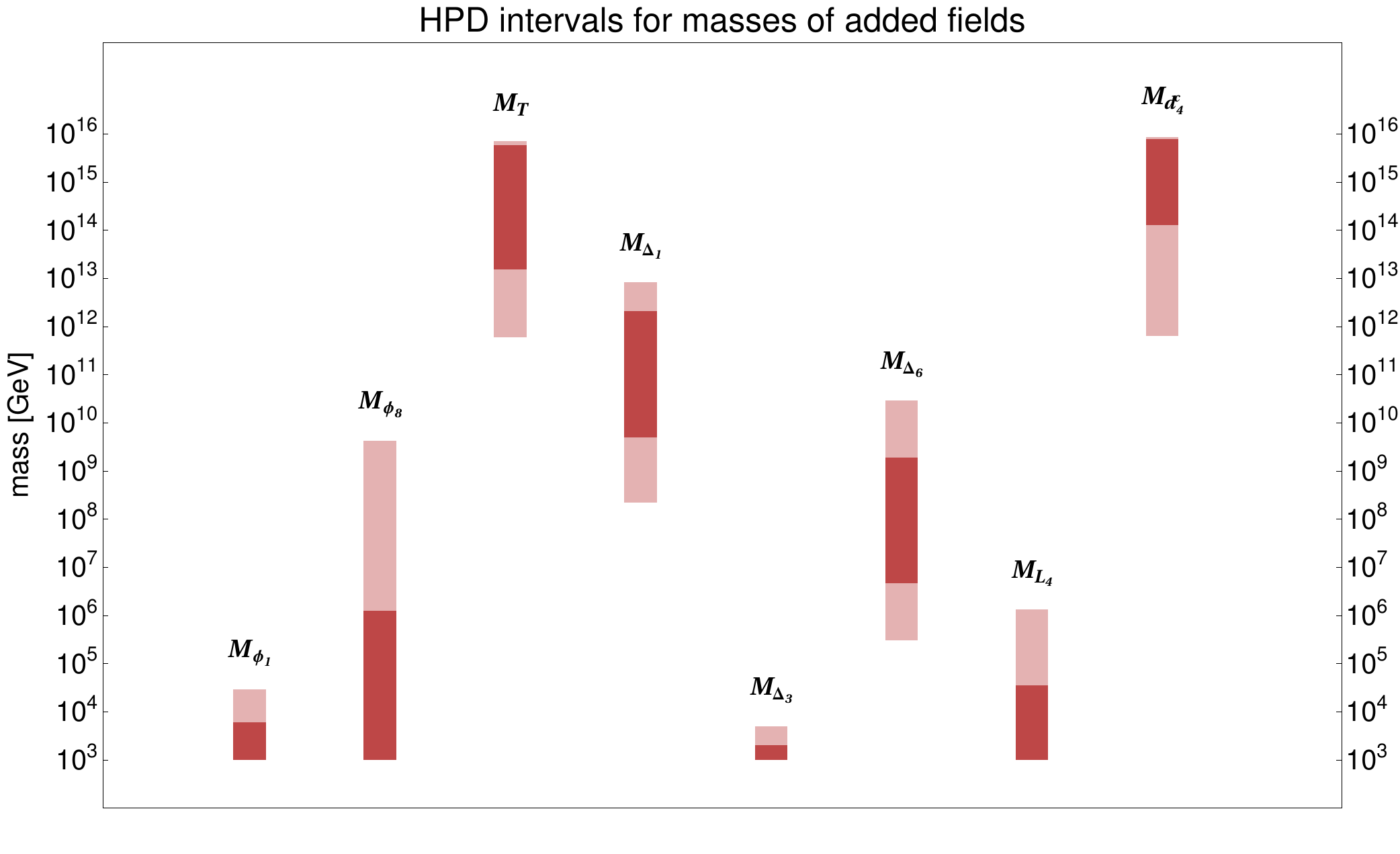}
    \caption{1$\sigma$ (dark) and 2$\sigma$ (light) HPD ranges of the masses of the added BSM states. }
    \label{fig:HPD masses}
    \vspace{15mm}
    \includegraphics[width=11cm]{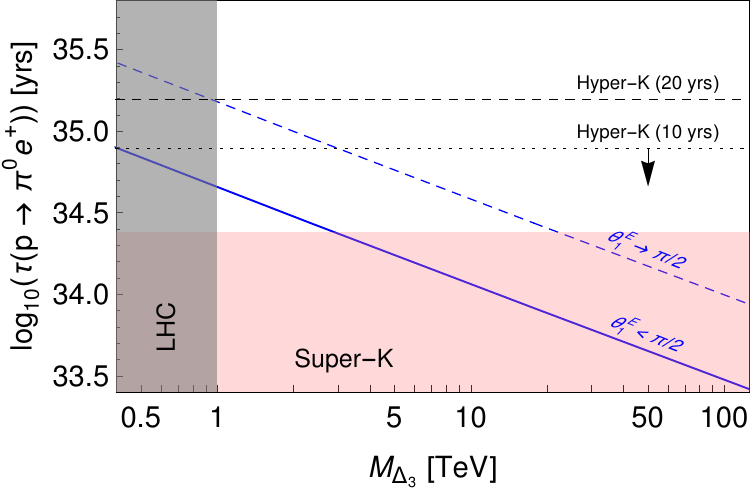}
    \caption{Dependence of the partial proton lifetime for the decay channel $p\rightarrow \pi^0e^+$ on the LQ mass. The upper bound of the partial proton lifetime is indicated by a solid blue line for a generic choice of the model parameter $\theta_1^E$. The dashed blue line corresponds to the special region in the parameter space, namely,  $\theta_1^E\rightarrow \pi/2$ (see text for details).  }
    \label{fig:LQ vs proton decay}
\end{figure}

From Eq.\ \eqref{eq:Gamma p->pi e}, the question arises whether the freedom in the flavor structure in the fermionic mass matrices can be used to rotate proton decay in the decay channel $p\rightarrow \pi^0 e^+$ away. Our findings show that it is indeed possible to suppress proton decay in this decay channel by an appropriate choice of the model parameters. However, it cannot be completely rotated away. Note that the decay channel under consideration is mostly dependent on the model parameter $\theta_1^E=\arctan(|M_1^E|/|M_4^E|)$. We present the dependence of the partial lifetime on $\theta_1^E$ for a benchmark scenario of gauge coupling unification in the left panel of Fig.\ \ref{fig:proton decay vs theta}. Clearly, proton decay gets suppressed by a bit more than half an order of magnitude for $\theta_1^{E,D}\rightarrow \pi/2$. The reason it cannot get suppressed further is that the decay width for this channel is obtained by the sum of two contributions, one of which is proportional to $c(e^c,d)$, while the other one is proportional to $c(e,d^c)$. Only, $c(e^c,d)$ depends on $E_R$ and $D_L$. Thus, only the contribution proportional to this c-factor (which is the dominant contribution for $\theta_1^E< \pi/2$) gets suppressed. Then, for $\theta_1^E$ close to $\pi/2$ the contribution proportional to the c-factor $c(e,d^c)$ becomes dominant and no further proton decay suppression is possible.

Another interesting result is the predicted range for the masses of the added fields. We present the $1\sigma$ (dark) and $2\sigma$ (light) HPD results for these masses in Fig.\ \ref{fig:HPD masses}. As discussed in Section\ \ref{sec:gauge coupling unification}, we vary all masses between the TeV scale and the GUT scale apart from $M_T$ which we keep above $3\times 10^{11}$\ GeV to sufficiently suppress scalar mediated proton decay. As already indicated by our gauge coupling unification plot (Fig.\ \ref{fig:gauge coupling unification}), there are four particles that can reside at the TeV scale, namely, the $SU(2)$ triplet $\phi_1$, the $SU(3)$ octet $\phi_8$, the LQ $\Delta_3$, and the VLD $L_4$. While $\phi_8$ and $L_4$ are also allowed to be heavier than 100 TeV, the upper bounds on the ranges for $\phi_1$ and $\Delta_3$ are relatively small. The upper bound of the $1\sigma$ ($2\sigma$) range for $M_{\phi_1}$ is 6\ TeV (29\ TeV). Moreover, for $M_{\Delta_3}$ we find an upper bound of the $1\sigma$ ($2\sigma$) range of 2\ TeV (5\ TeV). Since the predicted upper bound of the HPD range of the LQ mass is so small, we are further interested in its absolute bound and the correlation of this bound with the proton decay predictions. We therefore perform a fitting procedure maximizing the proton decay lifetime for a given constant LQ mass. The corresponding correlation between the LQ mass and the upper bound of the partial proton lifetime in the decay channel $p\rightarrow \pi^0 e^+$ is shown in Fig.\ \ref{fig:LQ vs proton decay}. The solid blue line shows the dependence of the maximal partial proton lifetime of the decay channel $p\rightarrow \pi^0e^+$ on the LQ mass $M_{\Delta_3}$ without using the freedom of the flavor structure of the fermion mass matrices to suppress proton decay, i.e., the case $\theta_1^E<\pi/2$. The dashed blue line shows the same relation where the flavor freedom is used to suppress proton decay, i.e., $\theta_1^E\rightarrow \pi/2$. If the flavor freedom is (not) used the current upper bound on the LQ mass is 20 TeV (3 TeV). In the future, for the case where the flavor freedom is used, this upper bound will be reduced to 3 TeV (900 GeV) if no proton decay is seen after 10 years (20 years) of runtime at Hyper-Kamiokande. Since the current LHC bound on this LQ mass is 1 TeV\ \cite{ATLAS:2019qpq}, intriguingly, Hyper-Kamiokande has the potential to fully test our model.

\begin{figure}
    \centering
    \includegraphics[width=\textwidth/2-3mm]{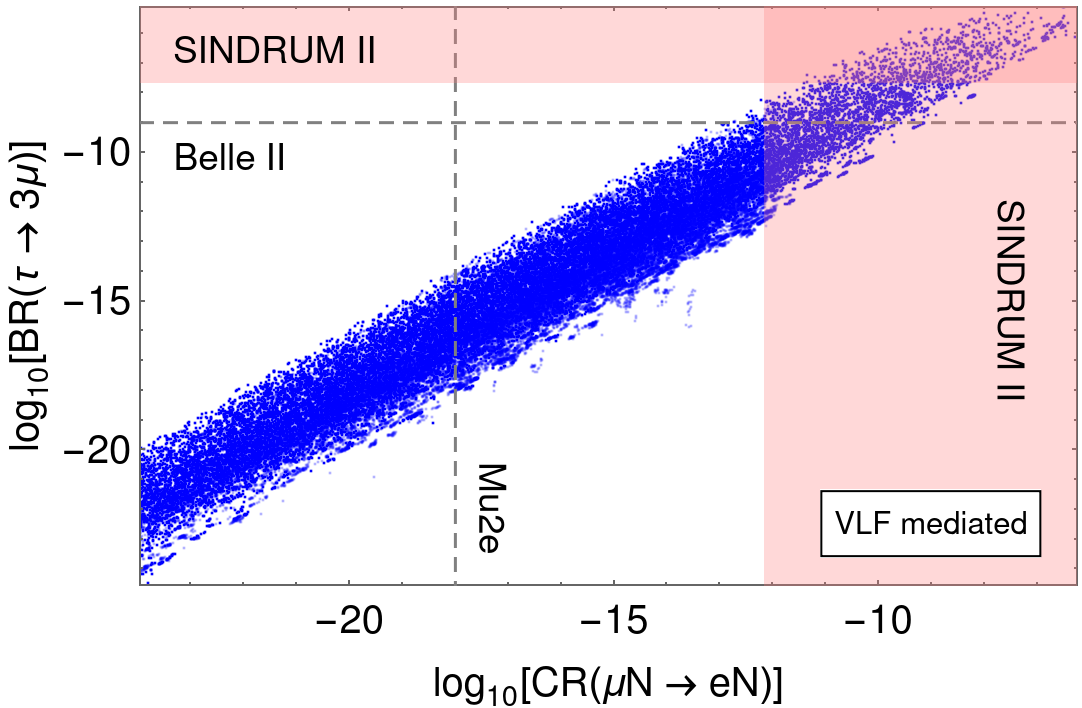}\hspace{3mm}
    \includegraphics[width=\textwidth/2-3mm]{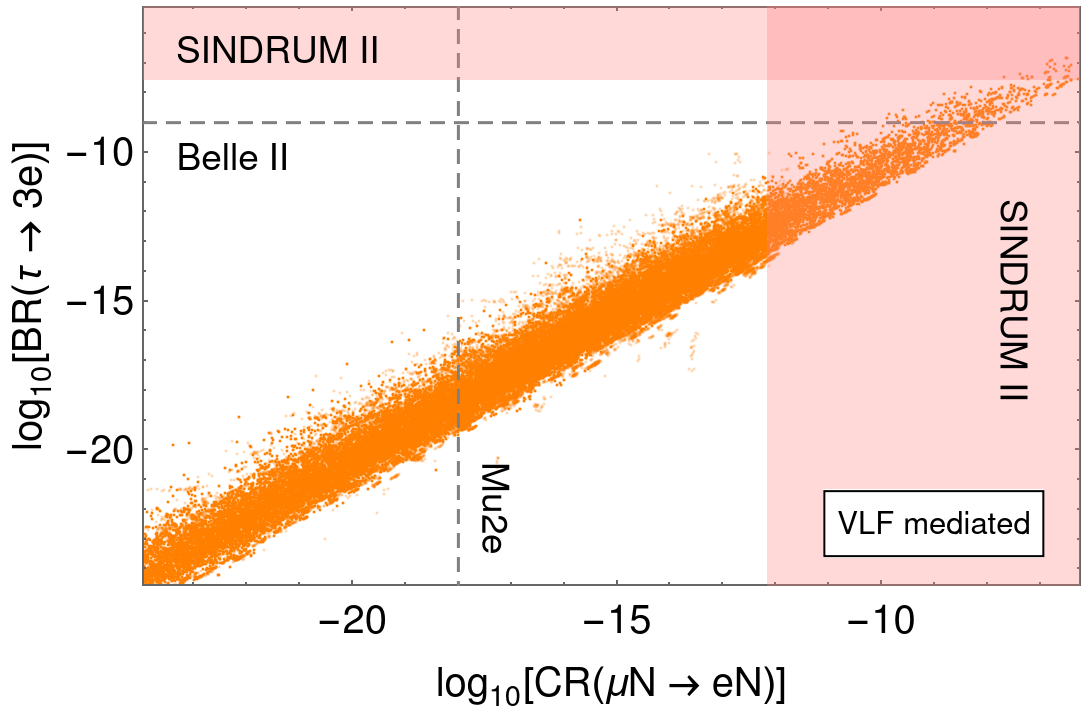}\\\vspace{5mm}
    \includegraphics[width=\textwidth/2-3mm]{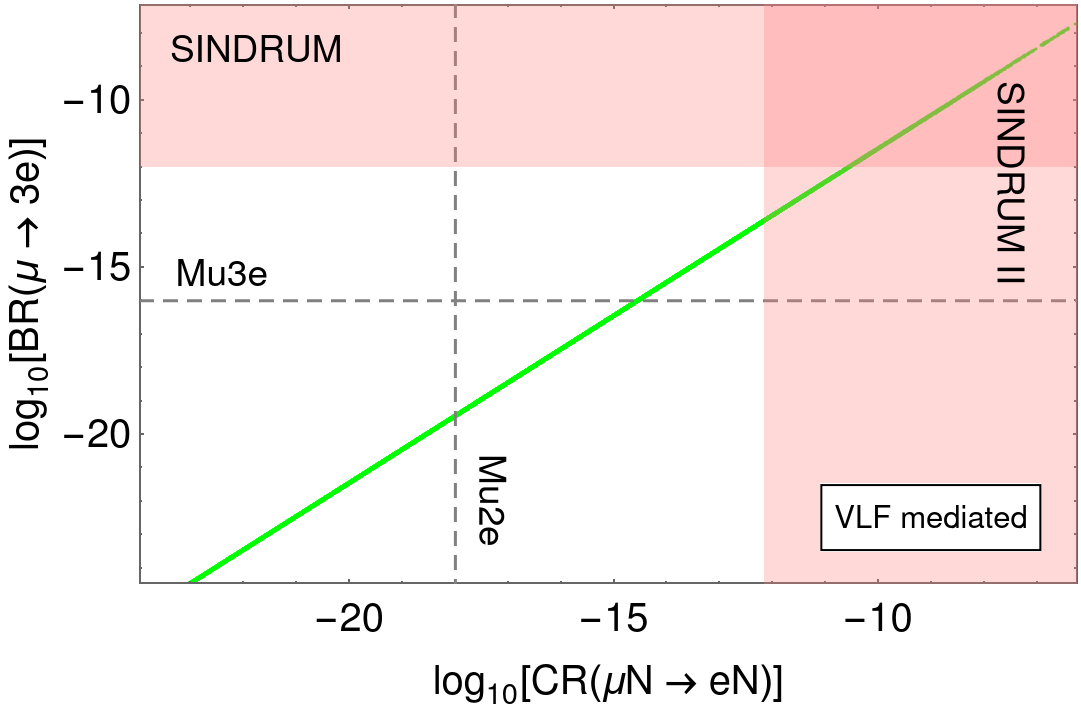}
    \caption{Relation between the flavor violating processes $\ell\rightarrow 3\ell^\prime$ and $\mu\rightarrow e$ conversion. }
    \label{fig:FV1}
\end{figure}

\begin{figure}
    \centering
    \includegraphics[width=\textwidth/2-3mm]{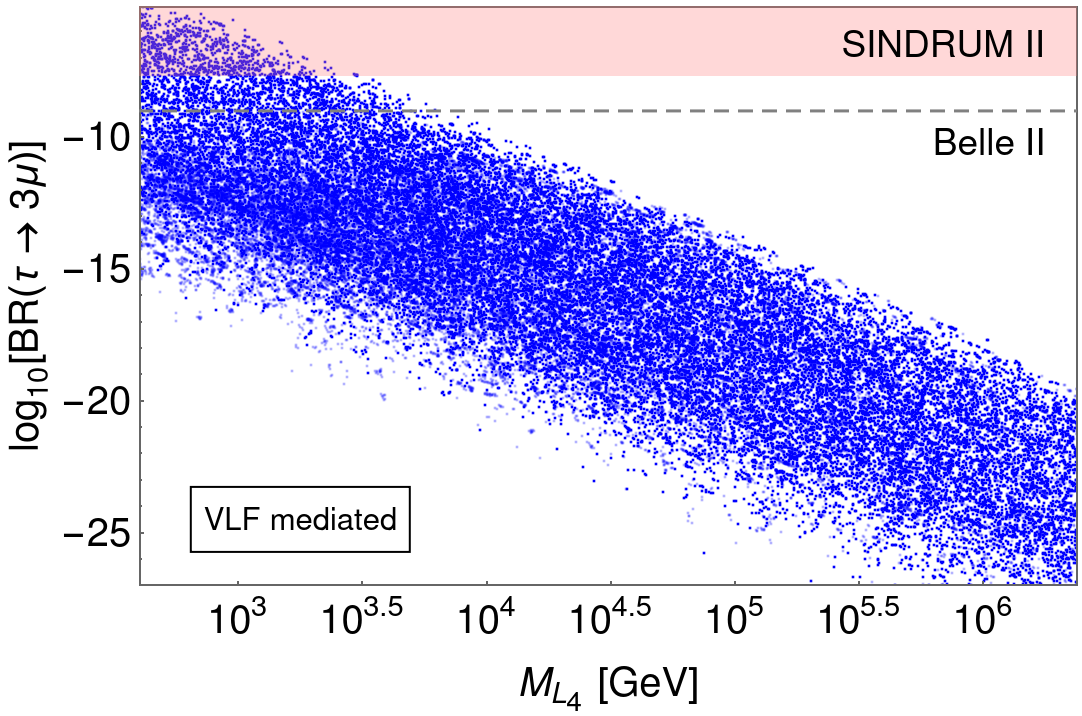}\hspace{3mm}
    \includegraphics[width=\textwidth/2-3mm]{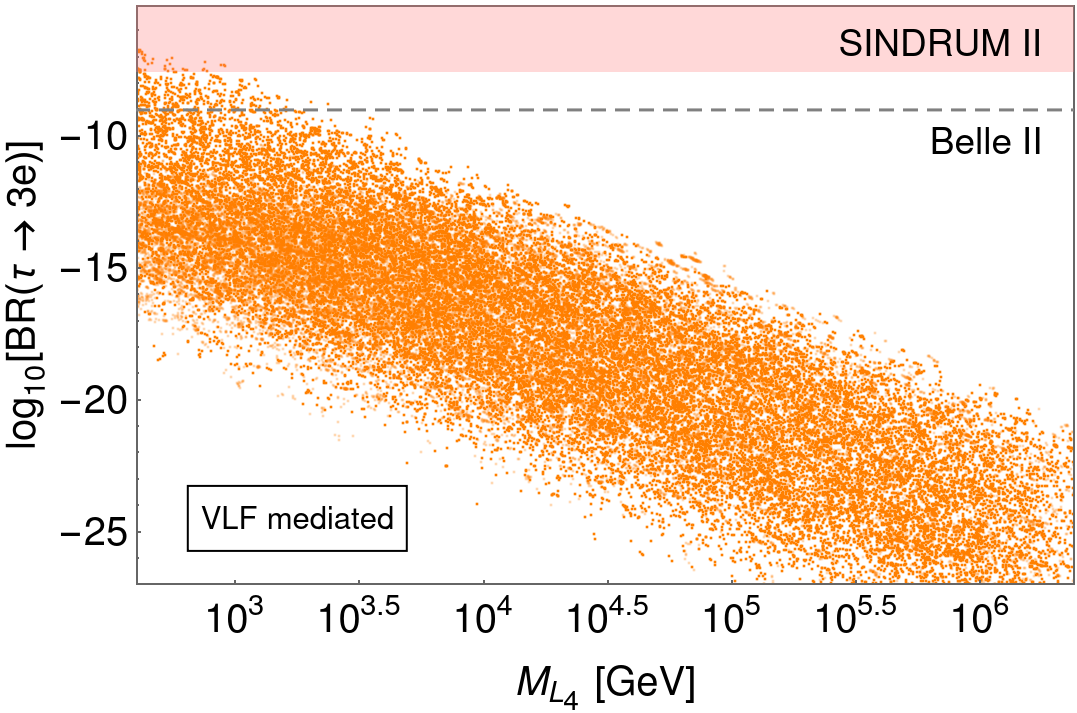}\\\vspace{5mm}
    \includegraphics[width=\textwidth/2-3mm]{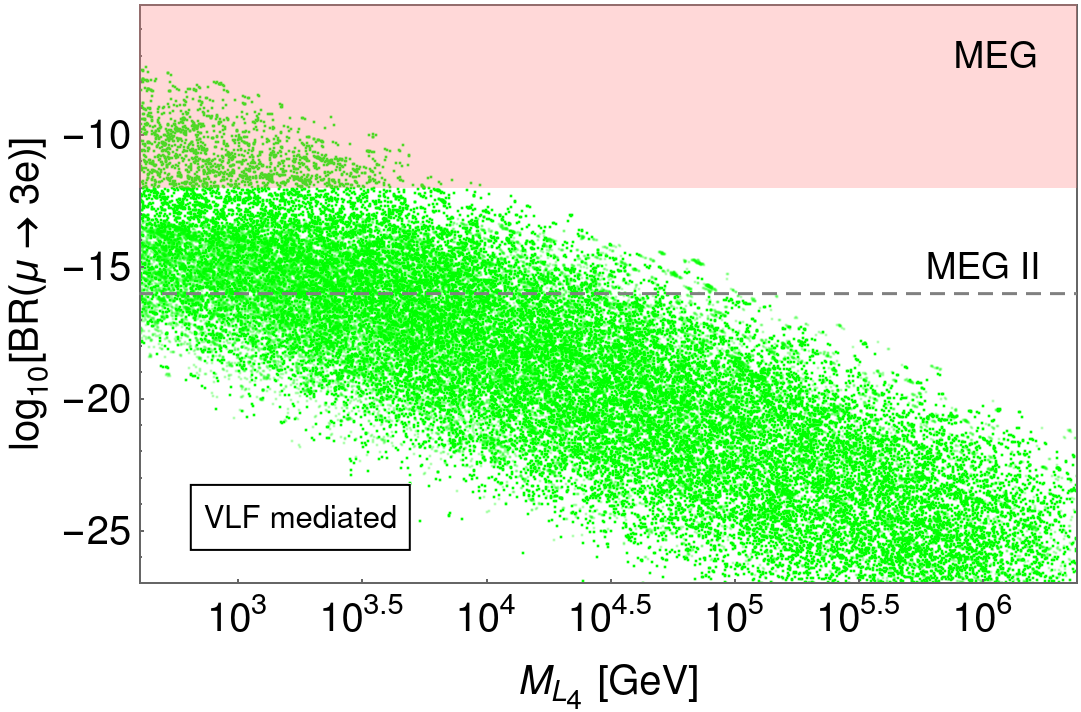}\hspace{3mm}
    \includegraphics[width=\textwidth/2-3mm]{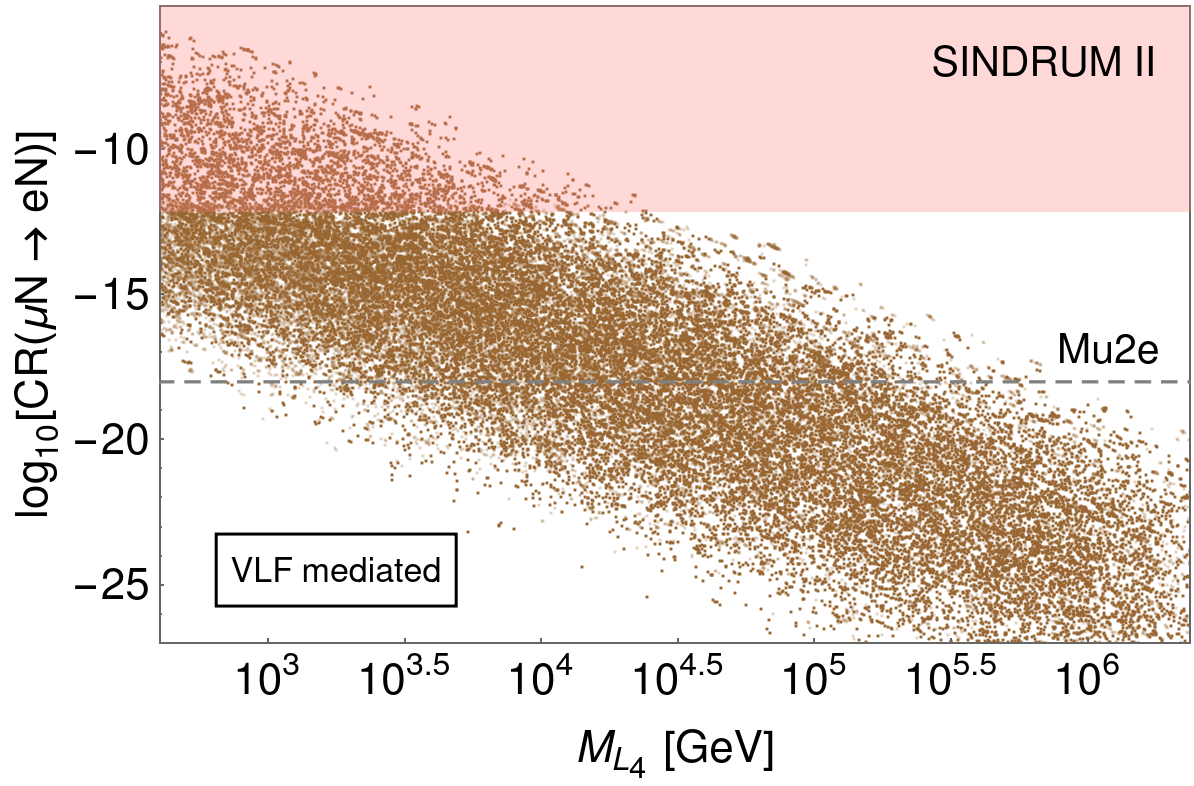}
    \caption{Relation between the VLD mass and the flavor violating processes $\ell\rightarrow 3 \ell^\prime$, respectively $\mu\rightarrow e$ conversion.  }
    \label{fig:FV2}
    \vspace{15mm}
    \includegraphics[width=\textwidth/2-3mm]{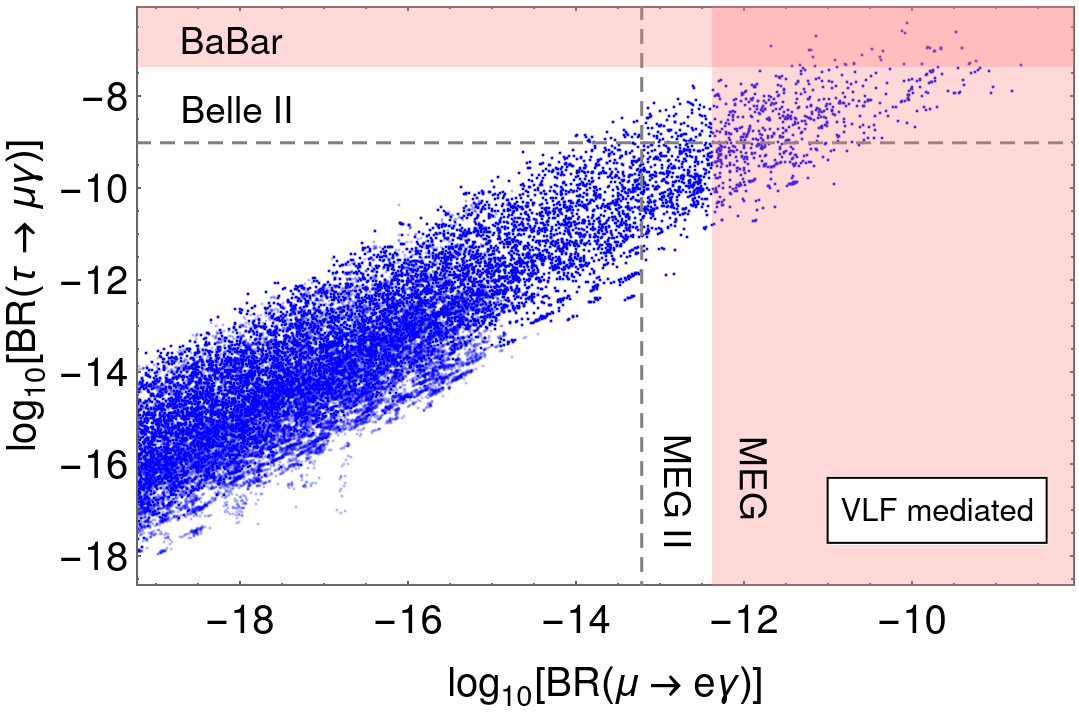}\hspace{3mm}
    \includegraphics[width=\textwidth/2-3mm]{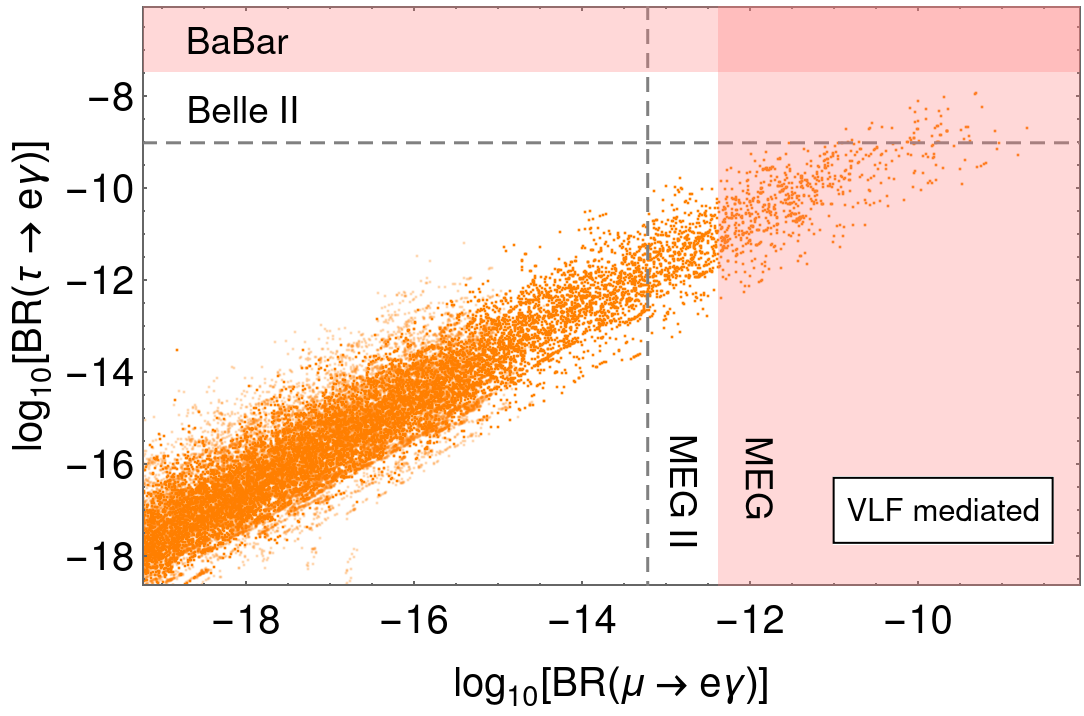}
    \caption{Correlation between the processes $\ell\rightarrow \ell^\prime \gamma$.}
    \label{fig:FV3}
\end{figure}

\begin{figure}[t!]
    \centering
    \includegraphics[width=\textwidth/2-3mm]{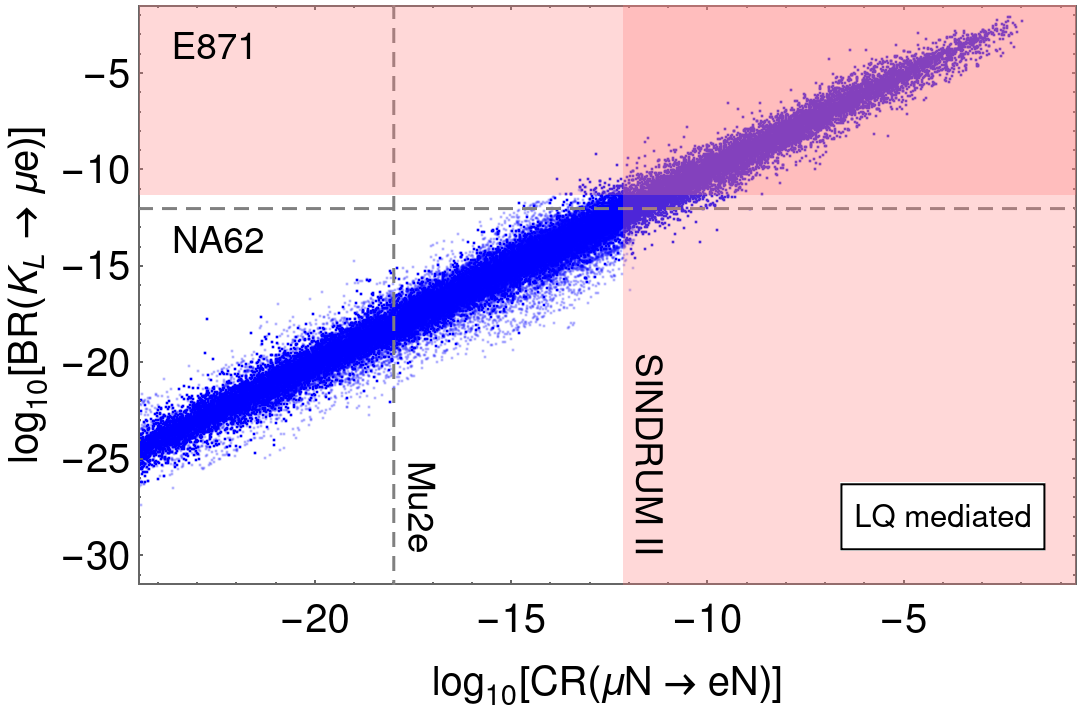}\hspace{3mm}
    \includegraphics[width=\textwidth/2-3mm]{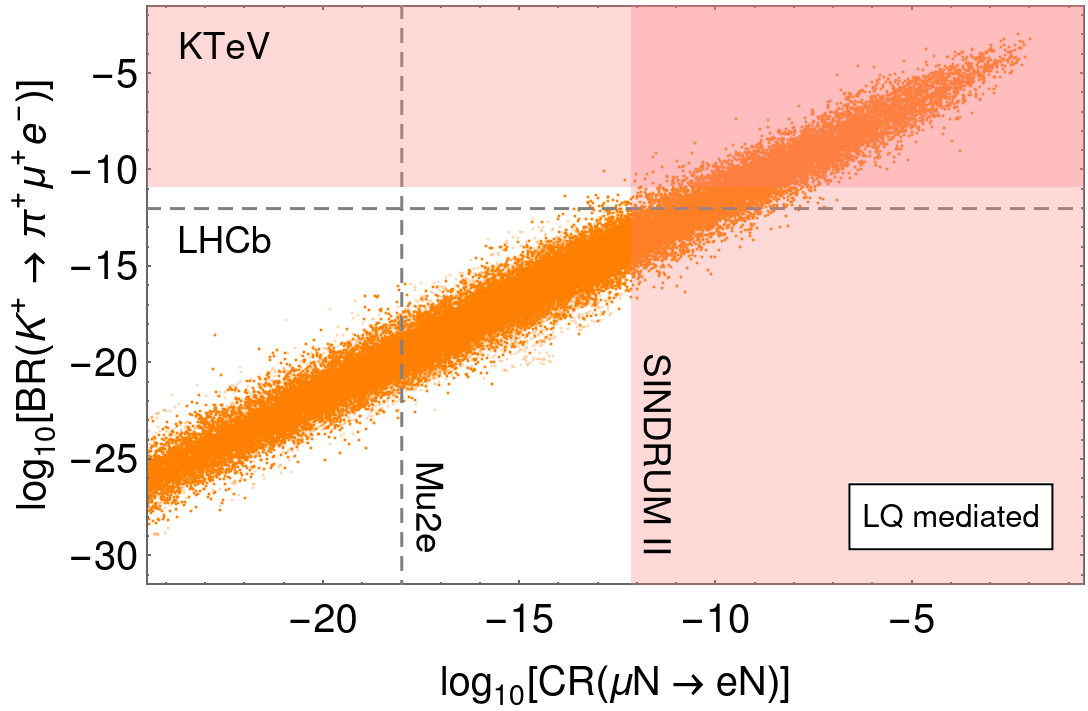}\\\vspace{5mm}
    \includegraphics[width=\textwidth/2-3mm]{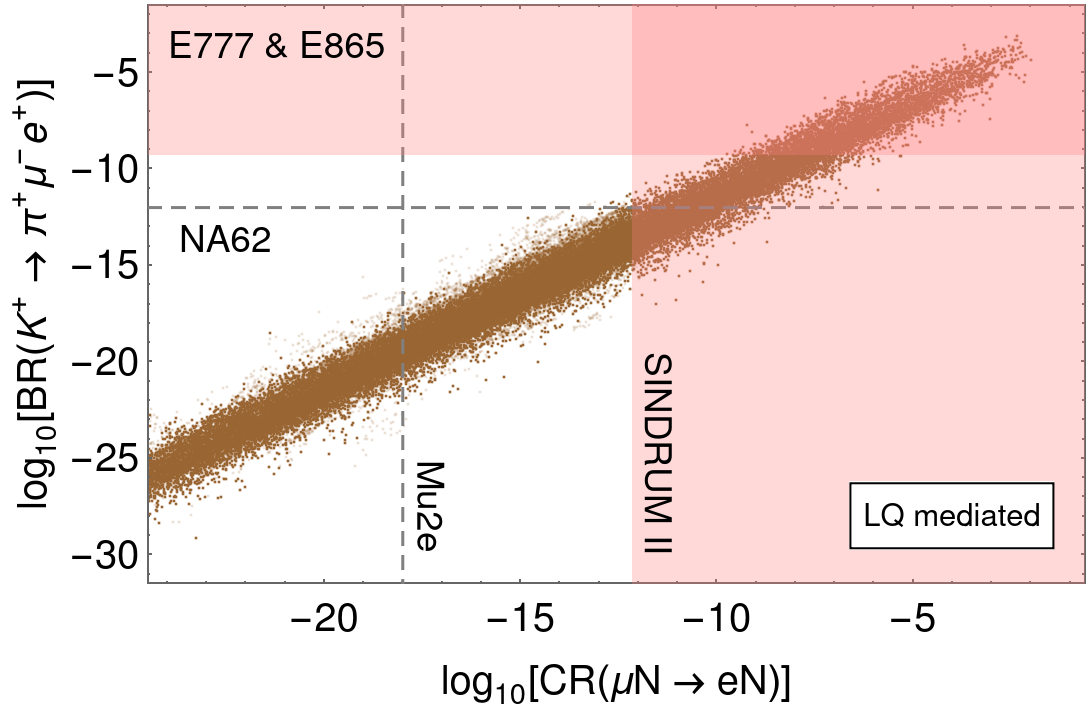}\hspace{3mm}
    \includegraphics[width=\textwidth/2-3mm]{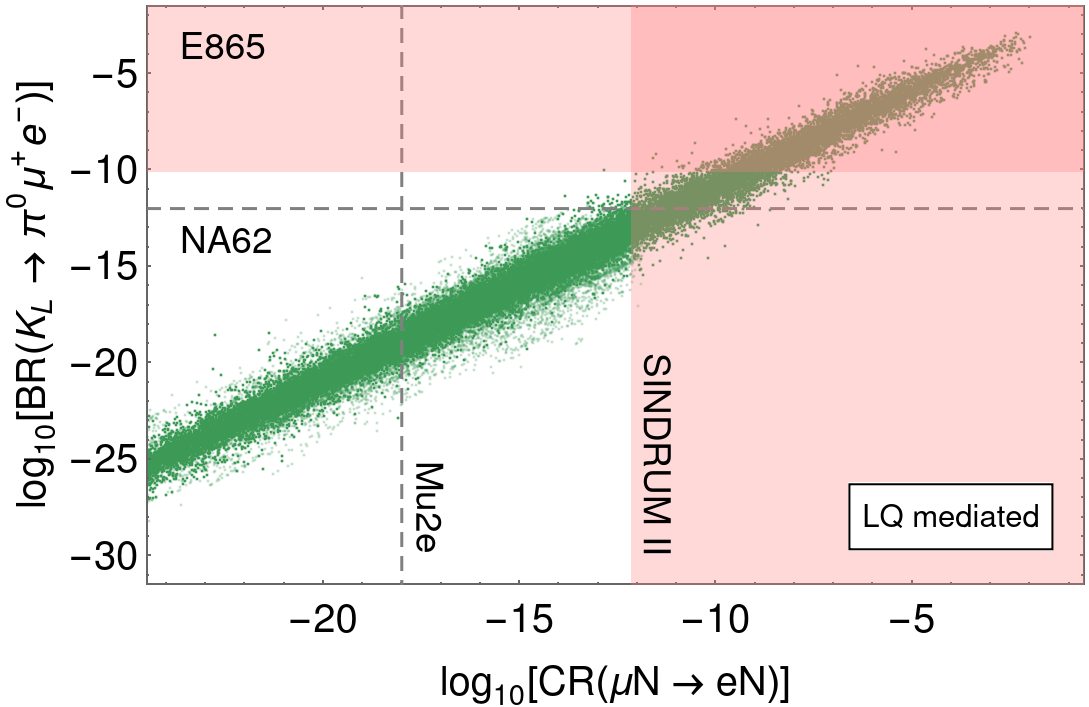}
    \caption{Correlation between various kaon decays and $\mu\rightarrow e$ conversion. }
    \label{fig:FV4}
\end{figure}

Other interesting predictions that can potentially be used to test our model are various flavor violating processes. Figures\ \ref{fig:FV1}-\ref{fig:FV4} show our predictions for some of these processes. We obtain the data that is visualized in these figures by performing MCMC analyses. The current experimental bounds are represented in all figures by the magenta colored regions, while future sensitivities of upcoming experiments are indicated with dashed lines. For all processes we assume normal neutrino mass ordering. We also analyze the case of inverted neutrino mass ordering, but it turns out, that the obtained relations are very similar to the normal ordering case. Therefore, we omit presenting the results obtained from the inverted neutrino ordering.

Figure\ \ref{fig:FV1} shows various relations between the processes that are mediated by a $Z$ boson exchange, namely, $\ell\rightarrow 3\ell^\prime$ and $\mu\rightarrow e$ conversion. The third panel is especially interesting since it suggests a strong correlation between the process $\mu\rightarrow 3e$ and $\mu\rightarrow e$ conversion.  From the first panel we deduce that if the process $\tau\rightarrow 3\mu$ is seen at the upcoming experiment, then we should also see a $\mu\rightarrow e$ conversion just above the current experimental constraint. Moreover, an observation of the process $\tau\rightarrow 3e$ at the upcoming experiment would highly disfavor our proposed model. Also, interesting correlations between these four processes with the VLD mass $M_{L_4}$ are depicted in Figure\ \ref{fig:FV2}.

The processes $\ell\rightarrow \ell^\prime\gamma$ stem from loop diagrams involving a $W$ or a $Z$ boson. The correlations between these processes are presented in Figure\ \ref{fig:FV3}. The left panel suggests that if the process $\tau\rightarrow \mu\gamma$ is observed, then $\mu\rightarrow e\gamma$ should also be seen. On the other hand, an observation of the process $\tau\rightarrow e\gamma$ would disfavor our model.

Various kaon decays as well as $\mu\rightarrow e$ conversion are mediated via an exchange of the LQ $\Delta_3$. The couplings of this LQ are related with the couplings for neutrino mass generation, since the LQ lives in the same $SU(5)$ representation as the weak triplet $\Delta_1$ that is responsible for neutrino mass generation. In Figure\ \ref{fig:FV4}, we show the correlation between $\mu\rightarrow e$ conversion and different kaon decays for a benchmark scenario, assuming that the two weak triplets $\Delta_1$ are mass degenerate and similarly the two LQs $\Delta_3$ are degenerate in mass (to maximize the GUT scale). The figure shows that although only a small part of the parameter space will be tested by upcoming experiments, there is still the potential to observe kaon decays. Such an observation would imply that $\mu\rightarrow e$ conversion should also be seen close to its current experimental bound.

Before concluding this section, we point out that in the supersymmetric (SUSY) framework (which we do not consider in this work), the wrong mass relations between the down-quark and charged lepton sectors can be resolved by the same mechanism as discussed above. The formulas of the Yukawa matrices derived at the GUT scale remain identical regardless of whether SUSY is imposed. For the mass matrices, in the case of SUSY,  $v_5\to v_u$ ($v_5\to v_d$) needs to be performed in the up-quark sector (in the down-quark and charged lepton sectors).  This is also true for the scenarios we explore in Sec.~\ref{sec:1015}. However, it is worth pointing out that in the case of SUSY with TeV scale sparticles, the minimal supersymmetric SM (MSSM) automatically guarantees gauge coupling unification close to $2\times 10^{16}$ GeV. Therefore, the phenomenological implications are completely different, since the low energy effective theory in the SUSY version is similar to the MSSM case. On the contrary, in the non-SUSY case studied in this paper, we find definite predictions that a specific set of beyond the SM states arising from GUT multiplets must be light close to the TeV scale to allow for gauge coupling unification and satisfy the proton decay bounds.

\section{Case study: $10_F+\overline{10}_F$/$15_F+\overline{15}_F$ VLFs}\label{sec:1015}
Instead of introducing VLFs in the fundamental representations, one can add VLFs in the  $10_F+\overline{10}_F$/$15_F+\overline{15}_F$ dimensional representations. 
In this section, we derive the full mass matrices of the fermions and discuss the gauge coupling unification in these scenarios.

\subsection{Case study: $10_F+\overline{10}_F$ VLFs}\label{sec:10}
As before, we introduce a single generation of VLF. We denote the component fields within 
 $10^a_F+\overline{10}_F$ as
\begin{align}
10^a_F=\frac{1}{\sqrt{2}} \begin{pmatrix}
0&u^c_b&-u^c_g&u_r&d_r\\
-u^c_g&0&u^c_1&u_2&d_g\\
u^c_g&-u^c_r&0&u_b&d_b\\
-u_r&-u_g&-u_b&0&e^c\\
-d_r&-d_g&-d_b&-e^c&0
\end{pmatrix}_a,\;\;\;
\overline{10}_F=\frac{1}{\sqrt{2}} \begin{pmatrix}
0&U_b&-U_g&U^c_r&D^c_r\\
-U_g&0&U_r&U^c_g&D^c_g\\
U_g&-U_r&0&U^c_b&D^c_b\\
-U^c_r&-U^c_g&-U^c_b&0&E^-\\
-D^c_r&-D^c_g&-D^c_b&-E^-&0
\end{pmatrix}\;.
\end{align}

With the addition of one generation of $10_F+\overline{10}_F$, the complete Yukawa part of the Lagrangian can be written as 
\begin{align}
&\mathcal{L}_Y= Y_{10}^{ab}10_F^a10^b_F5_H+Y_5^{ia} \overline 5^i_F 10^a_F 5^*_H   +y^\prime \overline{10}_F\overline{10}_F5^*_H + \left(m_a  + \lambda_a 24_H\right) \overline{10}_F  10_F^a.
\end{align}
Considering only the mass term for $10_F+\overline{10}_F$, i.e., the last term in the above equation by setting $a=4$, we find a mass relation among the submultiplet, which is given by
\begin{align}\label{eq:mass_relation_10F}
   M_{\widetilde Q}=\frac{1}{2} \left(M_{E^c}+M_{U^c}\right)\,.
\end{align}

From the above Yukawa interactions, it is straightforward to derive the fermion mass matrices, which we find to be
\begin{align}
\mathcal{L}_Y&\supset 
\begin{pmatrix}
d_1&d_2&d_3&d_4    
\end{pmatrix}
\begin{pmatrix}
\underbrace{\left(Y^T_5\right)^{ij}v_5}_{3\times 3}
&
\underbrace{m_i+\frac{\lambda_i v_{24}}{2}}_{3\times 1}  
\\
\underbrace{\left(Y^T_5\right)^{4j}v_5}_{1\times 3}
&
\underbrace{m_4+\frac{\lambda_4 v_{24}}{2}}_{1\times 1} 
\end{pmatrix}
\begin{pmatrix}
d_1^c\\d_2^c\\d_3^c\\D^c     
\end{pmatrix}
\\
&+\begin{pmatrix}
e_1&e_2&e_3&E^-    
\end{pmatrix}
\begin{pmatrix}
\underbrace{\left(Y_5\right)^{ij}v_5}_{3\times 3}
&
\underbrace{\left(Y_5\right)^{i4}v_5}_{3\times 1}
\\
\underbrace{m_j+ 3\lambda_j v_{24}}_{1\times 3}  
&
\underbrace{m_4+3\lambda_4 v_{24}}_{1\times 1} 
\end{pmatrix}
\begin{pmatrix}
e_1^c\\e_2^c\\e_3^c\\e_4^c     
\end{pmatrix}
\\
&+\begin{pmatrix}
u_1&u_2&u_3&u_4&U    
\end{pmatrix}
\begin{pmatrix}
\underbrace{4\left(Y_{10}+Y_{10}^T\right)^{ab}v_5}_{4\times 4}
&
\underbrace{m_a+\frac{\lambda_a v_{24}}{2}}_{4\times 1}
\\
\underbrace{m_b-2\lambda_b v_{24}}_{1\times 4}  
&
\underbrace{4y^\prime v_5}_{1\times 1} 
\end{pmatrix}
\begin{pmatrix}
u_1^c\\u_2^c\\u_3^c\\u_4^c\\U^c     
\end{pmatrix}\;.\label{eq:MU5x5}
\end{align}
As can be easily seen from these matrices, there are enough free parameters to correct the wrong mass relations between  the down-type quarks and the charged leptons. By integrating out the heavy states, the $3\times 3$ mass matrices of the corresponding light states can be written as 
\begin{align}
M_d=\left( \mathds{1}+ \frac{1}{|\eta_{Q4}|^2}\eta_Q\eta_Q^\dagger \right)^{-1/2} \left( v_5Y^T-\frac{v_5}{\eta_{Q4}}\eta_Q \hat Y \right),    
\end{align}
and 
\begin{align}
M_e=\left( v_5Y-\frac{v_5}{\eta_{L4}} \tilde Y \eta_L \right) \left( \mathds{1}+ \frac{1}{|\eta_{L4}|^2}\eta_L^\dagger\eta_L \right)^{-1/2} ,    
\end{align}
for the down-quark and charged lepton sectors, respectively, where we have defined the following quantities:
\begin{align}
&\eta_{Q4}=m_4+\frac{\lambda_4v_{24}}{2}, \;\;\;  
\eta_{L4}=m_4+3\lambda_4v_{24},
\\&
\eta_Q=\begin{pmatrix}
m_1+\frac{\lambda_1v_{24}}{2}\\
m_2+\frac{\lambda_2v_{24}}{2}\\
m_3+\frac{\lambda_3v_{24}}{2}
\end{pmatrix}, \;\;\;
\eta_L=\begin{pmatrix}
m_1+3\lambda_1v_{24}\\
m_2+3\lambda_2v_{24}\\
m_3+3\lambda_3v_{24}
\end{pmatrix}^T,
\\&
Y|_{3\times 3}=\left(Y_5\right)^{ij}, \;\;\;
\hat Y|_{1\times 3}=\left(Y_5^T\right)^{4j}, \;\;\;
\tilde Y|_{3\times 1}=\left(Y_5\right)^{i4}.
\end{align}
The $5\times 5$-dim up-type quark mass matrix, Eq.~\eqref{eq:MU5x5}, can be approximately block diagonalized as described in Appendix\ \ref{app:Block diagonalization of fermion mass matrices}. Afterwards, the $3\times 3$ block of light up-type quarks can be diagonalized using the usual numerical method.

\subsection{Case study: $15_F+\overline{15}_F$ VLFs}\label{sec:15}
In this case, we add one generation of $15_F+\overline{15}_F$. The decomposition of $15_F$ is as follows:
\begin{align}
    15_F=\Sigma_1(1,3,1)+\Sigma_3(3,2,1/6)+\Sigma_6(6,1,-2/3).
\end{align}
The Yukawa Lagrangian in the scenario takes the following form: 
\begin{align}
\mathcal{L}&\supset 
Y^u_{ij}\;10_{F\,i} 10_{F\,j}5_H
+Y^d_{ij}\;10_{F\,i}\overline{5}_{F\,j} 5^\ast_H
+Y^{a}_{i}\;15_F \overline{5}_{F\,i} 5^\ast_H
\nonumber \\ &
 +Y^{c}_{i}\; 10_{F\,i} \overline{15}_F 24_H
+\left(m_{15}+y\;24_H\right) \overline{15}_F 15_F 
+\mathrm{h.c.}\;.  
\end{align}
By considering only the last term (i.e., the $\overline{15}_F 15_F$ term), one obtains the following mass relation among the submultiplets:
\begin{equation}
\label{eq:mass_relation_15F}
M_{\Sigma_3}=\frac{1}{2} \left(M_{\Sigma_1}+M_{\Sigma_6}\right)\,.
\end{equation}

In this scenario, the mismatch between the down-type quarks and the charged leptons arises due to the mixing between the VLF and fermions in ${10_{F}}_i$. Once the GUT and the electroweak symmetries are broken,  the relevant decomposition under the $SU(3) \times U(1)_\mathrm{em}$ gauge group is $Q_i=u_i(3,2/3)+d_i(3,-1/3)$, $L_i=e_i(1,-1)+\nu_i(1,0)$, $\Sigma_3=\Sigma^u(3,2/3)+\Sigma^d(3,-1/3)$, and $\Sigma_1=\Sigma^{\nu}(1,0)+\Sigma^{e^c}(1,1)+ \Sigma^{e^ce^c}(1,2)$. Then the charged fermion mass matrices can be written as, 
\begin{align}
\mathcal{L} \supset &
\begin{pmatrix}u_i&\Sigma^u\end{pmatrix}
\begin{pmatrix}
\underbrace{4 v_5 Y^{U}}_{3\times 3} & \underbrace{\frac{5}{2}v_{24} Y^c}_{3\times 1} \\
\underbrace{0}_{1\times 3} &\underbrace{M_{\Sigma_3}}_{1\times 1}
\end{pmatrix}
\begin{pmatrix}u^c_j\\\overline{\Sigma}^u\end{pmatrix}   
\\& 
+\begin{pmatrix}d_i&\Sigma^d\end{pmatrix}
\begin{pmatrix}
\underbrace{v_5 Y^d}_{3\times 3} & \underbrace{\frac{5}{2}v_{24} Y^c}_{3\times 1}\\
\underbrace{v_5 Y^a}_{1\times 3} & \underbrace{M_{\Sigma_3}}_{1\times 1}
\end{pmatrix}
\begin{pmatrix}d^c_j\\\overline{\Sigma}^d\end{pmatrix}
+\begin{pmatrix}e_i&\overline{\Sigma}^{e^c}\end{pmatrix}
\begin{pmatrix}
\underbrace{v_5 \left(Y^d\right)^T}_{3\times 3} & \underbrace{v_5 Y^a}_{3\times 1}\\ \underbrace{0}_{1\times 3} &\underbrace{M_{\Sigma_1}}_{1\times 1}
\end{pmatrix}
\begin{pmatrix}e^c_j\\\Sigma^{e^c}\end{pmatrix}\,, \nonumber 
\end{align}
where we have defined $Y^{U}=Y^u+\left(Y^u\right)^T$.

Finally, integrating out the heavy states, we can write the $3\times 3$ mass matrices of the light SM fermions as, 
\begin{align}
&M_u=\left(\mathds{1}+\frac{25v^2_{24}}{4M^2_{\Sigma_3}} \;Y^c{Y^c}^{\dagger} \right)^{-\frac{1}{2}} 4 v_5 Y^{U}, 
\\
&M_d=\left(\mathds{1}+\frac{25v^2_{24}}{4M^2_{\Sigma_3}}\;Y^c{Y^c}^{\dagger} \right)^{-\frac{1}{2}} v_5 \left( Y^d -\frac{5v_{24}}{2M_{\Sigma_3}}\; Y^cY^a  \right),  
\\
&M_e=v_5 \left(Y^d\right)^T.
\end{align}
This case also provides a sufficient number of free parameters to correct the bad mass relations~\cite{Dorsner:2019vgf,Dorsner:2021qwg,Antusch:2023jok}. See also Refs.\ \cite{Shafi:1999rm, Shafi:1999ft,Oshimo:2009ia} for SUSY models with vectorlike 15-plets.

\subsection{Gauge coupling unification}
\begin{figure}[t]
    \centering
    \includegraphics[width=0.55\textwidth]{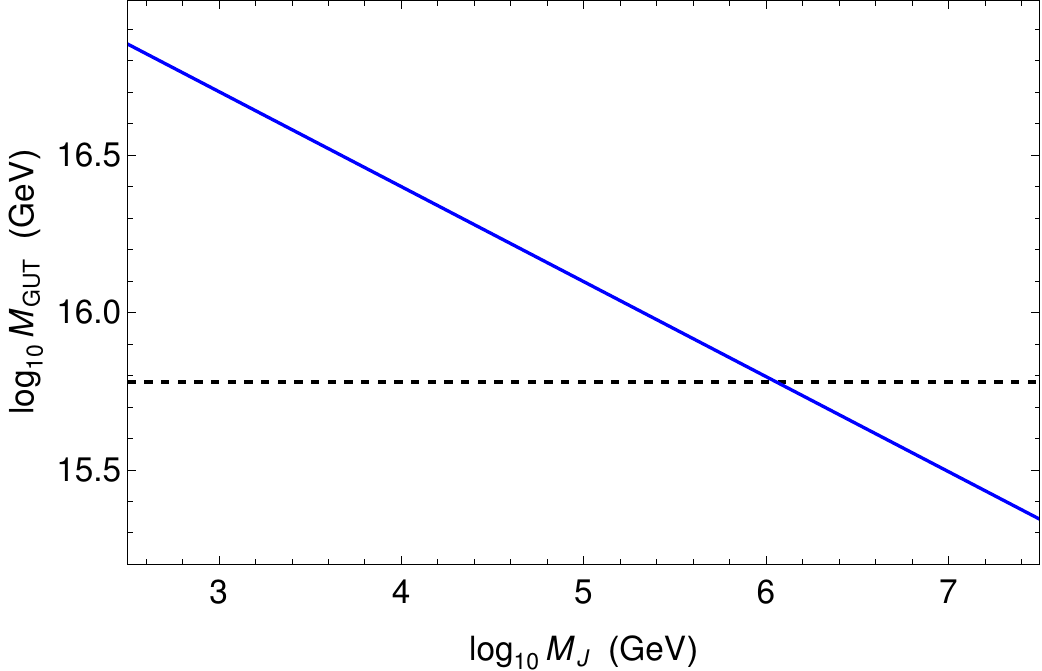}
    \caption{The maximal GUT scale as a function of the smallest mass of intermediate-scale particles for the case where a VLF in the representation $10_F+\overline{10}_F$ or $15_F+\overline{15}_F$ has been added to the GG model. See text for details. }
    \label{fig:max_MGUT_10F15F}
\end{figure}

The two-loop beta function for the gauge coupling unification can be found in Section\ \ref{sec:gauge coupling unification}, and the relevant one-loop and two-loop gauge coefficients are listed in Appendix\ \ref{app:RGE}. Both $SU(5)$ representations, $10_F$ and $15_F$, contain a multiplet $(3,2,1/6)$, which can mix with the SM left-chiral quark doublet, i.e. $10_F\supset \widetilde Q(3,2,1/6)$ and $15_F\supset \Sigma_3(3,2,1/6)$. In both cases, the GUT scale is maximized if this fermionic multiplet $\widetilde Q$ ($\Sigma_3$) is kept light together with the weak triplet $\phi_1$ and color octet $\phi_8$ of the adjoint Higgs field, while the remaining multiplets in $10_F$ ($15_F$) and $24_H$ reside at the GUT scale. This choice of masses automatically respects the mass relations provided in Eq.\ \eqref{eq:mass_relation_10F}, respectively Eq.\ \eqref{eq:mass_relation_15F}. Figure\ \ref{fig:max_MGUT_10F15F} shows the maximal GUT scale as a function of the intermediate mass scale $M_J$, which refers to the mass $M_{\widetilde Q}$ or $M_{\Sigma_3}$, while $M_{\Phi_1}$ and $M_{\Phi_8}$ are varied between $M_J$ and $M_{\textrm{GUT}}$. The horizontal dashed line approximately indicates the GUT scale ($M_\mathrm{GUT}\sim 6\times 10^{15}$ GeV) that is required to evade the current proton decay bound without using the flavor freedom of the fermion mass matrices. In the case of $10_F+\overline{10}_F$, utilizing additional freedom from the flavor sector, proton decay can be further suppressed such that even lower GUT scales become viable. This freedom in the flavor sector 
does not exist in the case of $15_F+\overline{15}_F$ as discussed in Refs.\ \cite{Dorsner:2019vgf, Dorsner:2021qwg, Antusch:2023jok}.

\section{Conclusions}\label{sec:con}
This work aimed to determine the minimal viable renormalizable $SU(5)$ GUT with representations no higher than adjoints. We concluded that an $SU(5)$ model containing a pair of VLFs $5_F+\overline{5}_F$ and two copies of $15_H$ Higgs fields serves as the minimal candidate, satisfying the requirements for correct charged fermion and neutrino masses while addressing the matter-antimatter asymmetry of the universe. Our findings demonstrate that this proposed model possesses a high degree of predictability, and will undergo comprehensive testing through a combination of upcoming proton decay experiments, collider searches, and low-energy experiments targeting flavor violation. Additionally, we explored the possibility of incorporating either $10_F+\overline{10}_F$ or $15_F+\overline{15}_F$ VLFs instead of $5_F+\overline{5}_F$ to correct the wrong mass relations. However, our study revealed that the entire parameter space of these alternative models, even with minimal particle content, cannot be fully probed by the next round of experiments due to the potential for long proton lifetimes that lie beyond the capabilities of Hyper-Kamiokande.

\subsection*{Acknowledgments}
S.S\; would like to thank Ilja Dor\v{s}ner for discussion.

\appendix
\section{Gauge coefficients of added multiplets}\label{app:RGE}
The RG running of the SM gauge couplings depends on the gauge coefficients of the intermediate scale fields. The one-loop gauge coefficients read
\begin{align}
    &
    a_i^{\phi_8}
    =\begin{pmatrix}
        0 & 0 & \frac{1}{2}
    \end{pmatrix},
    &&
    a_i^{\phi_1}=
    \begin{pmatrix}
        0 & \frac{1}{3} & 0
    \end{pmatrix},
    &&
    a_i^{L_4}=
    \begin{pmatrix}
        \frac{1}{5} & \frac{1}{3} & 0
    \end{pmatrix},
    &&
    a_i^{\overline{L}_4}=
    \begin{pmatrix}
        \frac{1}{5} & \frac{1}{3} & 0
    \end{pmatrix},
    \nonumber\\
    &
    a_i^{d^c_4}=
    \begin{pmatrix}
        \frac{2}{15} & 0 & \frac{1}{3}
    \end{pmatrix},
    &&
    a_i^{\overline{d^c_4}}=
    \begin{pmatrix}
        \frac{2}{15} & 0 & \frac{1}{3}
    \end{pmatrix},
    &&
    a_i^T=
    \begin{pmatrix}
        \frac{1}{15} & 0 & \frac{1}{6}
    \end{pmatrix},
    &&
    a_i^{\Delta_1}=
    \begin{pmatrix}
        \frac{3}{5} & \frac{2}{3} & 0
    \end{pmatrix},
    \nonumber\\
    &
    a_i^{\Delta_3}=
    \begin{pmatrix}
        \frac{1}{30} & \frac{1}{2} & \frac{1}{3}
    \end{pmatrix},
    &&
    a_i^{\Delta_6}=
    \begin{pmatrix}
        \frac{8}{15} & 0 & \frac{5}{6}
    \end{pmatrix}, 
    &&
    a_i^{\Sigma_1}=
    \begin{pmatrix}
        \frac{6}{5} & \frac{4}{3} & 0
    \end{pmatrix},
    &&
    a_i^{\overline{\Sigma}_1}=
    \begin{pmatrix}
        \frac{6}{5} & \frac{4}{3} & 0
    \end{pmatrix},
    \nonumber\\
    &
    a_i^{\Sigma_3}=
    \begin{pmatrix}
        \frac{1}{15} & 1 & \frac{2}{3}
    \end{pmatrix},
    &&
    a_i^{\overline{\Sigma}_3}=
    \begin{pmatrix}
        \frac{1}{15} & 1 & \frac{2}{3}
    \end{pmatrix},
    &&
    a_i^{\Sigma_6}=
    \begin{pmatrix}
        \frac{16}{15} & 0 & \frac{5}{3}
    \end{pmatrix},
    &&
    a_i^{\overline{\Sigma}_6}=
    \begin{pmatrix}
        \frac{16}{15} & 0 & \frac{5}{3}
    \end{pmatrix}.
\end{align}
The two-loop gauge coefficients are given by
\begin{align}
    &
    b_{ij}^{\phi_8}=
    \begin{pmatrix}
        0 & 0 & 0 \\
        0 & 0 & 0 \\
        0 & 0 & 21
    \end{pmatrix},
    &&
    b_{ij}^{\phi_3}=
    \begin{pmatrix}
        0 & 0 & 0 \\
        0 & \frac{28}{3} & 0 \\
        0 & 0 & 0
    \end{pmatrix},
    &&
    b_{ij}^{L_4}=
    \begin{pmatrix}
        \frac{9}{100} & \frac{9}{20} & 0 \\
        \frac{3}{20} & \frac{49}{12} & 0 \\
        0 & 0 & 0
    \end{pmatrix},
    &&
    b_{ij}^{\overline{L}_4}=
    \begin{pmatrix}
        \frac{9}{100} & \frac{9}{20} & 0 \\
        \frac{3}{20} & \frac{49}{12} & 0 \\
        0 & 0 & 0
    \end{pmatrix},
    \nonumber\\
    &
    b_{ij}^{d^c_4}=
    \begin{pmatrix}
        \frac{2}{75} & 0 & \frac{8}{15} \\
        0 & 0 & 0 \\
        \frac{1}{15} & 0 & \frac{19}{3}
    \end{pmatrix}
    &&
    b_{ij}^{\overline{d^c_4}}=
    \begin{pmatrix}
        \frac{2}{75} & 0 & \frac{8}{15} \\
        0 & 0 & 0 \\
        \frac{1}{15} & 0 & \frac{19}{3}
    \end{pmatrix},
    &&
    b_{ij}^T=
    \begin{pmatrix}
        \frac{4}{75} & 0 & \frac{16}{15} \\
        0 & 0 & 0 \\
        \frac{2}{15} & 0 & \frac{11}{3}
    \end{pmatrix},
    &&
    b_{ij}^{\Delta_1}=
    \begin{pmatrix}
        \frac{108}{25} & \frac{72}{5} & 0 \\
        \frac{24}{5} & \frac{56}{3} & 0 \\
        0 & 0 & 0
    \end{pmatrix},
    \nonumber\\
    & 
    b_{ij}^{\Delta_3}=
    \begin{pmatrix}
        \frac{1}{150} & \frac{3}{10} & \frac{8}{15} \\
        \frac{1}{10} & \frac{13}{2} & 8 \\
        \frac{1}{15} & 3 & \frac{22}{3}
    \end{pmatrix},
    &&
    b_{ij}^{\Delta_6}=
    \begin{pmatrix}
        \frac{128}{75} & 0 & \frac{64}{3} \\
        0 & 0 & 0 \\
        \frac{8}{3} & 0 & \frac{115}{3}
    \end{pmatrix},
    &&
    b_{ij}^{\Sigma_1}=
    \begin{pmatrix}
        \frac{54}{25} & \frac{36}{5} & 0 \\
        \frac{12}{5} & \frac{64}{3} & 0 \\
        0 & 0 & 0 
    \end{pmatrix},
    &&
    b_{ij}^{\overline{\Sigma}_1}=
    \begin{pmatrix}
        \frac{54}{25} & \frac{36}{5} & 0 \\
        \frac{12}{5} & \frac{64}{3} & 0 \\
        0 & 0 & 0 
    \end{pmatrix},
    \nonumber\\
    &
     b_{ij}^{\Sigma_3}=
    \begin{pmatrix}
        \frac{1}{300} & \frac{3}{20} & \frac{4}{15} \\
        \frac{1}{20} & \frac{49}{4} & 4 \\
        \frac{1}{30} & \frac{3}{2} & \frac{38}{3} 
    \end{pmatrix},
    &&
    b_{ij}^{\overline{\Sigma}_3}=
    \begin{pmatrix}
        \frac{1}{300} & \frac{3}{20} & \frac{4}{15} \\
        \frac{1}{20} & \frac{49}{4} & 4 \\
        \frac{1}{30} & \frac{3}{2} & \frac{38}{3} 
    \end{pmatrix},
    &&
     b_{ij}^{\Sigma_6}=
    \begin{pmatrix}
        \frac{64}{75} & 0 & \frac{32}{3} \\
        0 & 0 & 0 \\
        \frac{4}{3} & 0 & \frac{125}{3}
    \end{pmatrix},
    &&
    b_{ij}^{\overline{\Sigma}_6}=
    \begin{pmatrix}
        \frac{64}{75} & 0 & \frac{32}{3} \\
        0 & 0 & 0 \\
        \frac{4}{3} & 0 & \frac{125}{3}
    \end{pmatrix}.
\end{align}
\section{Block diagonalization of fermion mass matrices}\label{app:Block diagonalization of fermion mass matrices}
Defining the transformation matrices 
\begin{align}
    &P_L^E=\text{diag}(e^{-i\arg M_1^E},e^{-i\arg M_2^E},e^{-i\arg M_3^E},e^{-i\arg M_4^E}),
    \\
    &P_R^E=\text{diag}(e^{i\arg (M_1^E/m_1)},e^{i\arg (M_2^E/m_2)},e^{i\arg (M_3^E/m_3)},1),
    \\
    &P_L^D=\text{diag}(e^{i\arg (M_1^D/m_1)},e^{i\arg (M_2^D/m_2)},e^{i\arg (M_3^D/m_3)},1),
    \\
    &P_R^D=\text{diag}(e^{-i\arg M_1^D},e^{-i\arg M_2^D},e^{-i\arg M_3^D},e^{-i\arg M_4^D}),
    \\
    &V_{E,D}=
    \begin{pmatrix}
    c_1^{E,D} & -s_1^{E,D}s_2^{E,D} & -c_2^{E,D}s_1^{E,D}s_3^{E,D} & c_2^{E,D}c_3^{E,D}s_1^{E,D}
    \\
    0 & c_2^{E,D} & - s_2^{E,D}s_3^{E,D} & c_3^{E,D}s_2^{E,D} 
    \\
    0 & 0 & c_3^{E,D} & s_3^{E,D}
    \\
    -s_1^{E,D} & -c_1^{E,D}s_2^{E,D} & -c_1^{E,D}c_2^{E,D}s_3^{E,D} & c_1^{E,D}c_2^{E,D}c_3^{E,D}
    \end{pmatrix},
\end{align}
where $c_i^{E,D}=\cos\theta_i^{E,D}$ and $s_i^{E,D}=\sin\theta_i^{E,D}$, $i=1,2,3$, and where the angles $\theta_i^{E,D}$ are given by
\begin{align}\label{eq:relation theta_i^{D,E} - M_i^{D,E}}
    \tan\theta_1^{E,D}=\frac{|M_1^E|}{|M_4^E|},\qquad 
    \tan\theta_2^{E,D}=\frac{|M_2^E|}{|M_4^E|}\cos\theta_1^{E,D},\qquad
    \tan\theta_3^{E,D}=\frac{|M_3^E|}{|M_4^E|}\cos\theta_1^{E,D}\cos\theta_2^{E,D},
\end{align}
we can approximately block diagonalize the $4\times 4$ down-type and charged lepton mass matrices via
\begin{align}
    &M_D^\text{bd}= P_L^D M_D P_R^D V_D =
    \begin{pmatrix}
        m_1 c_1^D & -m_1 s_1^D s_2^D & -m_1 c_2^D s_1^D s_3^D & m_1 c_2^D c_3^D s_1^D
        \\
        0 & m_2 c_2^D & -m_2 s_2^D s_3^D & m_2 c_3^D s_2^D 
        \\
        0 & 0 & m_3 c_3^D & m_3 s_3^D
        \\
        0 & 0 & 0 & M_{d_4^c}
    \end{pmatrix},
    \\
    &M_E^\text{bd}=V_E^T P_L^E M_E P_R^E=
    \begin{pmatrix}
        m_1 c_1^E & 0 & 0 & 0 
        \\
        -m_1 s_1^E s_2^E & m_2 c_2^E & 0 & 0
        \\
        -m_1 c_2^E s_1^E s_3^E & -m_2 s_2^E s_3^E & m_3 c_3^E & 0 
        \\
        m_1 c_2^E c_3^E s_1^E & m_2 c_3^E s_2^E & m_3 s_3^E & M_{L_4}
    \end{pmatrix}.
\end{align}
In the case in which the VLQ (VLL) mass is much larger than $m_i$, the matrix $M_D^\text{bd}$ ($M_E^\text{bd}$) is approximately block diagonalized. Otherwise, the block diagonal form can only be achieved after using an additional left (right) rotation matrix corrections with angles of the order $m_i/M_{d_4^c}$ ($m_i/M_{L_4}$).

With a similar strategy the $5\times 5$ neutrino mass matrix can be approximately block diagonalized. Denoting $c_i^{N}=\cos\theta_i^{N}$ and $s_i^{N}=\sin\theta_i^{N}$, with $i\in \{1,2,3\}$, we define the angles $\theta_i^{N}$ as 
\begin{align}
    \tan\theta_1^N=-\frac{|M_1^E|}{|M_2^E|},\qquad 
    \tan\theta_2^N=-\frac{\sqrt{|M_1^E|^2+|M_2^E|^2}}{|M_3^E|},\qquad
    \tan\theta_3^N=-\frac{\sqrt{|M_1^E|^2+|M_2^E|^2+|M_3^E|^2}}{|M_4^E|}.
\end{align}
Using the transformation matrices
\begin{align}
    P_L^N&=\text{diag}(e^{-i\arg M_1^E},e^{-i\arg M_2^E},e^{-i\arg M_3^E},e^{-i\arg M_4^E},1),
    \\
    V_N&=
    \begin{pmatrix}
        c_1^N & -c_2^Ns_1^N & c_3^Ns_1^Ns_2^N & -\frac{1}{\sqrt{2}}s_1^Ns_2^Ns_3^N & \frac{i}{\sqrt{2}}s_1^Ns_2^Ns_3^N
        \\
        s_1^N & c_1^Nc_2^N & -c_1^Nc_3^Ns_2^N & \frac{1}{\sqrt{2}}c_1^Ns_2^Ns_3^N & -\frac{i}{\sqrt{2}}c_1^Ns_2^Ns_3^N
        \\
        0 & s_2^N & c_2^Nc_3^N & -\frac{1}{\sqrt{2}}c_2^Ns_3^N & \frac{i}{\sqrt{2}}c_2^Ns_3^N 
        \\
        0 & 0 & s_3^N & \frac{1}{\sqrt{2}}c_3^N & -\frac{i}{\sqrt{2}}c_3^N 
        \\
        0 & 0 & 0 & \frac{1}{\sqrt{2}} & \frac{i}{\sqrt{2}}
    \end{pmatrix}
\end{align}
allows for an approximate block diagonalization of the neutrino mass matrix via
\begin{align}
    M_N^\text{bd}=V_N^T P_L^N M_N {P_L^N} V_N=
    \begin{pmatrix}
        M_N^{3\times 3} & \mathcal{O}(\textrm{eV}) & \mathcal{O}(\textrm{eV})\\
        \mathcal{O}(\textrm{eV}) & M_{L_4} & \mathcal{O}(\textrm{eV}) \\
        \mathcal{O}(\textrm{eV}) & \mathcal{O}(\textrm{eV}) & M_{L_4}
    \end{pmatrix},
\end{align}
The off-diagonal blocks in $M_N^\text{bd}$ are of the same order as the neutrino masses, i.e.\ sub eV, and thus much smaller than the VLD mass.

Similarly, the $5\times 5$-dim up-type quark mass matrix $M_U^{5\times 5}$ that is obtained by adding a pair of vectorlike fermionic 10-plets (see Eq.\ \eqref{eq:MU5x5} in Section\ \ref{sec:10}) can be approximately block diagonalized via
\begin{align}
    M_U^{\textrm{bd}}={V_U^L}^TP_U^LM_U^{5\times 5}P_U^RV_U^R=
    \begin{pmatrix}
        M_U^{3\times 3} & \mathcal{O}(m_t) & \mathcal{O}(m_t)\\
        \mathcal{O}(m_t) & M_Q & \mathcal{O}(m_t) \\
        \mathcal{O}(m_t) & \mathcal{O}(m_t) & M_{u^c_4}
    \end{pmatrix}.
\end{align}
Here, we have defined the transformation matrices
\begin{align}
    &P_L^U=\text{diag}(e^{-i\arg \eta_{Q1}},e^{-i\arg \eta_{Q2}},e^{-i\arg \eta_{Q3}},e^{-i\arg \eta_{Q4}},1),\\
    &P_R^U=\text{diag}(e^{-i\arg \eta_{U1}},e^{-i\arg \eta_{U2}},e^{-i\arg \eta_{U3}},e^{-i\arg \eta_{U4}},1),\\
    &V_L^U=
    \begin{pmatrix}
        c_1^{U,L} & -c_2^{U,L}s_1^{U,L} & c_3^{U,L}s_1^{U,L}s_2^{U,L} & -s_1^{U,L}s_2^{U,L}s_3^{U,L} & 0 \\
        s_1^{U,L} & c_1^{U,L}c_2^{U,L} & - c_1^{U,L}c_3^{U,L}s_2^{U,L} & c_1^{U,L}s_2^{U,L}s_3^{U,L} & 0 \\
        0 & s_2^{U,L} & c_2^{U,L}c_3^{U,L} & -c_2^{U,L}s_3^{U,L} & 0 \\
        0 & 0 & s_3^{U,L} & c_3^{U,L} & 0 \\
        0 & 0 & 0 & 0 & 1
    \end{pmatrix},\\
    &V_R^U=
    \begin{pmatrix}
        c_1^{U,R} & -c_2^{U,R}s_1^{U,R} & c_3^{U,R}s_1^{U,R}s_2^{U,R} & 0 & -s_1^{U,R}s_2^{U,R}s_3^{U,R} \\
        s_1^{U,R} & c_1^{U,R}c_2^{U,R} & - c_1^{U,R}c_3^{U,R}s_2^{U,R} & 0 & c_1^{U,R}s_2^{U,R}s_3^{U,R} \\
        0 & s_2^{U,R} & c_2^{U,R}c_3^{U,R} & 0 & -c_2^{U,R}s_3^{U,R} \\
        0 & 0 & s_3^{U,R} & 0 & c_3^{U,R} \\
        0 & 0 & 0 & 1 & 0
    \end{pmatrix},
\end{align}
using the quantities
\begin{align}
    &\eta_{Ua}=m_a-2-\lambda _a v_{24},\\
    &M_{\tilde{Q}}=\sqrt{\eta_{Q_1}^2+\eta_{Q_2}^2+\eta_{Q_3}^2+\eta_{Q_4}^2},\qquad
    M_{u^c_4}=\sqrt{\eta_{U_1}^2+\eta_{U_2}^2+\eta_{U_3}^2+\eta_{U_4}^2},\\
    &t_1^{U,L}=-\frac{|\eta_{Q1}|}{|\eta_{Q2}|},\qquad 
    t_2^{U,L}=-\frac{\sqrt{|\eta_{Q1}|^2+|\eta_{Q2}|^2}}{|\eta_{Q3}|},\qquad
    t_3^{U,L}=-\frac{\sqrt{|\eta_{Q1}|^2+|\eta_{Q2}|^2+|\eta_{Q3}|^2}}{|\eta_{Q4}|},\\
    &t_1^{U,R}=-\frac{|\eta_{U1}|}{|\eta_{U2}|},\qquad 
    t_2^{U,R}=-\frac{\sqrt{|\eta_{U1}|^2+|\eta_{U2}|^2}}{|\eta_{U3}|},\qquad 
    t_3^{U,R}=-\frac{\sqrt{|\eta_{U1}|^2+|\eta_{U2}|^2+|\eta_{U3}|^2}}{|\eta_{U4}|},
\end{align}
and applying the notation $s_i^{U,L/R}=\sin\theta_i^{U,L/R}$, $c_i^{U,L/R}=\cos\theta_i^{U,L/R}$, and $t_i^{U,L/R}=\tan\theta_i^{U,L/R}$, with $i\in\lbrace 1,2,3\rbrace$. 

\bibliographystyle{style}
\bibliography{references}
\end{document}